\documentclass[journal,10pt]{IEEEtran}
\usepackage{algorithm}
\usepackage{algpseudocode}
\usepackage{amsfonts}
\usepackage{amsmath}
\usepackage{amssymb}
\usepackage{bm}
\usepackage{breqn}
\usepackage[font=small,labelsep=period]{caption}
\usepackage{cases}
\usepackage{cite}
\usepackage{color}
\usepackage{enumerate}
\usepackage{epsfig}
\usepackage{epstopdf}
\usepackage{exscale}
\usepackage{float}
\usepackage{graphics}
\usepackage{graphicx}
\usepackage{latexsym}
\usepackage{multicol}
\usepackage{multirow}
\usepackage{relsize}
\usepackage{setspace}
\usepackage{stfloats}
\usepackage{subfigure}
\usepackage{url}

\begin{document}


\title{Performance Analysis and Optimization for RIS-Assisted Multi-User Massive MIMO Systems with Imperfect Hardware}
\IEEEoverridecommandlockouts


\author{Zhangjie~Peng,
        Xianzhe~Chen,
        Cunhua~Pan,~\IEEEmembership{Member,~IEEE},
        Maged~Elkashlan,~\IEEEmembership{Member,~IEEE},
        and~Jiangzhou~Wang,~\IEEEmembership{Fellow,~IEEE}

\thanks{Copyright (c) 2015 IEEE. Personal use of this material is permitted. However, permission to use this material for any other purposes must be obtained from the IEEE by sending a request to pubs-permissions@ieee.org.}
\thanks{This work was supported in part by the Natural Science Foundation of Shanghai under Grant 22ZR1445600, in part by the open research fund of National Mobile Communications Research Laboratory, Southeast University under Grant 2018D14, and in part by the National Natural Science Foundation of China under Grant 61701307. \emph{(Corresponding author: Cunhua Pan, Xianzhe Chen)}.}
\thanks{Zhangjie Peng is with the College of Information, Mechanical and Electrical Engineering, Shanghai Normal University, Shanghai 200234, China, also with the National Mobile Communications Research Laboratory, Southeast University, Nanjing 210096, China, and also with the Shanghai Engineering Research Center of Intelligent Education and Bigdata, Shanghai Normal University, Shanghai 200234, China (e-mail: pengzhangjie@shnu.edu.cn).}
\thanks{Xianzhe Chen is with the College of Information, Mechanical and Electrical Engineering, Shanghai Normal University, Shanghai 200234, China, and also with the Department of Electrical and Computer Engineering, The University of British Columbia, Vancouver, BC V6T1Z4, Canada (e-mail: cxzdubu001@163.com).}
\thanks{Cunhua Pan is with the National Mobile Communications Research Laboratory, Southeast University, Nanjing 210096, China (e-mail: cpan@seu.edu.cn).}
\thanks{Maged Elkashlan is with the School of Electronic Engineering and Computer Science at Queen Mary University of London, London E1 4NS, U.K. (e-mail: maged.elkashlan@qmul.ac.uk).}
\thanks{Jiangzhou Wang is with the School of Engineering, University of Kent, Canterbury, CT2 7NT, United Kingdom (e-mail: j.z.wang@kent.ac.uk).}
}

\maketitle

\newtheorem{lemma}{Lemma}
\newtheorem{theorem}{Theorem}
\newtheorem{remark}{Remark}
\newtheorem{corollary}{Corollary}
\newtheorem{proposition}{Proposition}

\vspace{-1cm}
\begin{abstract}
The paper studies a reconfigurable intelligent surface (RIS)-assisted multi-user uplink massive multiple-input multiple-output (MIMO) system with imperfect hardware.
At the RIS, the paper considers phase noise, while at the base station, the paper takes into consideration the radio frequency impairments and low-resolution analog-to-digital converters.
The paper derives approximate expressions for the ergodic achievable rate in closed forms under Rician fading channels.
For the cases of infinite numbers of antennas and infinite numbers of reflecting elements, asymptotic data rates are derived to provide new design insights.
The derived power scaling laws indicate that while guaranteeing a required system performance, the transmit power of the users can be scaled down at most by the factor $\frac{1}{M}$ when $M$ goes infinite, or by the factor $\frac{1}{MN}$ when $M$ and $N$ go infinite, where $M$ is the number of antennas and $N$ is the number of the reflecting units.
Furthermore, an optimization algorithm is proposed based on the genetic algorithm to solve the phase shift optimization problem with the aim of maximizing the sum rate of the system.
Additionally, the optimization problem with discrete phase shifts is considered.
Finally, numerical results are provided to validate the correctness of the analytical results.

\end{abstract}


\section{Introduction}
As fifth generation (5G) commercial networks began to be deployed in 2020, there is an increasing demand for the future communication systems to support the ever-increasing number of devices while guaranteeing the high quality of the communication service \cite{8663968}.
Massive multiple-input multiple-output (MIMO) has been widely used to enhance the system performance, since it can efficiently reduce multi-user interference and scale down the transmit power of users \cite{6832435,9137716,9400853}.
Recently, reconfigurable intelligent surface (RIS), also known as intelligent reflecting surface (IRS) or large intelligent surface (LIS), has been proposed as a promising technique to extend the coverage and improve the spectrum efficiency (SE) and energy efficiency (EE) of communication networks \cite{8319526,9140329,8936989,8941126,9475160,9328501}.
RIS is mainly composed of numerous reflecting elements, each of which imposes an independent phase shift on the incident signals.
In particular, by carefully tuning the phase shifts, RIS can shape the wireless radio propagation environment to be customized to meet specific targets.
Different from the relay, the reflecting elements at RIS require no active hardware components such as radio frequency (RF) chains and amplifiers, which significantly reduce the power consumption and hardware cost.
Furthermore, RIS can work in the full-dulplex mode without the self-loop interference \cite{9318531}.

Due to its appealing advantages, RIS has attracted extensive research attention from both academia and industry.
For instance, the authors in \cite{8746155} studied an RIS-assisted single-user downlink system, derived the upper bound expression of the SE, and optimized the phase shifts at the RIS.
In the case where the working base station (BS) was interfered by another BS, the authors in \cite{9184252} obtained tractable expressions for the data rates under both instantaneous and statistical channel state information (CSI).
In \cite{9099236}, the authors applied RIS to Internet of Things, analyzed the system performance and proposed a time-length allocation scheme for minimizing the energy consumption.
The authors in \cite{9181610} investigated a multi-RIS downlink system with imperfect location information of the users, and analyzed the impacts of the system parameters with derived approximate expressions for the ergodic AR.
In \cite{9366346}, the authors studied a multi-pair system assisted by RIS, and utilized the genetic algorithm (GA) to solve the phase shift optimization problem for maximizing the sum ergodic achievable rate (AR).
In \cite{9387559}, the authors in took into account the interplay of the responses between the phase shift and the amplitude in an RIS-assisted single-user system.
The authors in \cite{9392378} investigated a downlink RIS-assisted massive MIMO system with statistical CSI at the BS, and optimized the beamformings at the RIS and the BS.
The authors in \cite{9530675} derived closed-form sum rates for RIS-assisted uplink systems under spatially correlated Rician Fading, and jointly optimized the phase-shifting matrix and the transmit covariance matrix.
In \cite{9743440}, the authors focused on an RIS-assisted multi-user uplink massive MIMO system, and proposed a GA-based algorithm for maximizing the sum ergodic AR according to the derived closed-form expressions.

It should be noted that the assumption of perfect hardware was considered in the aforementioned works.
However, in practice, communication systems suffer from imperfect hardware, which leads to hardware impairments (HIs), such as oscillator phase noise, in-phase/quadrature-phase (I/Q) imbalance, non-linearities and quantization errors \cite{7106472,7208847}.
Although compensation algorithms can alleviate the effects of HIs \cite{B_RFim,5456453,7080890}, the residual HIs still degrade and limit the system performance.
Therefore, it is necessary to take HIs into consideration when analyzing practical systems.
The authors in \cite{9239335} considered HIs at the transmitters in an RIS-assisted single-user downlink system. They first obtained closed-form solutions for the optimal beamforming at the multi-antenna source, and then optimized the phase shifts at the RIS for maximizing the signal-to-noise ratio (SNR).
In \cite{9374557}, the secrecy rate was studied in RIS-assisted systems with HIs at the transmitters of the BS. The authors proposed an iterative method to optimize the beamforming vectors at both the BS and the RIS.
For a multi-RIS-assisted full-duplex system, the authors in \cite{9394419} investigated the impacts of HIs at the transceivers of the BS and users, and jointly optimized the beamformings at the BS and the RISs and the transmit power of the users for maximizing the sum AR.

Moreover, in massive MIMO systems, the BS is equipped with a large number of antennas, which requires a large number of analog-to-digital converters/digital-to-analog converters (ADCs/DACs), leading to a high power consumption and hardware cost. To address this, researchers use low-resolution ADCs/DACs to reduce the power consumption and hardware cost while sacrificing a certain system performance \cite{7979627,9075256,8629287}.
The authors in \cite{9295369} investigated an RIS-assisted single-user uplink system with low-resolution ADCs at the BS, and derived the AR expressions and analyzed the system performance.
In \cite{9483943}, an RIS-assisted multi-user downlink system with low-resolution DACs was studied. The authors derived the approximate expressions for AR and optimized the phase shifts at the RIS.

On the other hand, the phase noise has recently gain much attention in RIS-assisted systems, which is generated from the non-ideal reconfiguration of the phase shifts at the RIS in practical systems \cite{8869792}.
Considering phase noise at the RIS, the authors in \cite{9219155} studied an RIS-assisted system with two legitimate nodes and one eavesdropper, and analyzed the system performance based on the secrecy rate.
The authors in \cite{9322510} considered RIS-assisted single-user systems with phase noise and transceiver HIs, and analyzed the EE under Rayleigh fading channels.
The authors in \cite{9417454} extended the work to imperfect CSI cases, and studied the power scaling laws.
In \cite{9079457}, the authors investigated the SE and EE of an RIS-assisted downlink system with phase noise and RF impairments, assuming determined line-of-sight (LoS) channels.
The authors in \cite{9390410} derived the closed-form expressions for the AR of an RIS-assisted downlink system with phase noise and transceiver HIs, and optimized the phase shifts for maximizing the SNR.
The authors in \cite{9477418} optimized the transmit power, the beamformings at the BS and the RIS of an RIS-assisted multi-user system with phase noise and transceiver HIs under correlated Rayleigh fading channels and instantaneous CSI.
In \cite{9534477}, assuming correlated Rayleigh fading channels and statistical CSI, the authors studied the channel estimation, and optimized the phase shifts of RIS-assisted multi-user systems with phase noise and transceiver HIs.

In this paper, we focus on an RIS-assisted multi-user uplink massive MIMO system under Rician fading channels and with imperfect hardware.
We jointly consider the phase noise at the RIS, and the RF impairments and low-resolution ADCs at the BS.
Based on that, we derive the closed-form expressions for the ergodic AR, analyze the system performance and optimize the phase shifts.
The main contributions of this paper are summarized as follows:
\begin{itemize}
  \item We investigate an RIS-assisted multi-user uplink massive MIMO system under Rician fading channels and with imperfect hardware. The channels between the users and the RIS and that between the RIS and the BS are modelled as Rician fading. Furthermore, we jointly consider the phase noise at the RIS, and the RF impairments and low-resolution ADCs at the BS, due to the imperfect hardware.
  \item We derive approximate expressions for the ergodic AR in closed forms. Based on that, asymptotic data rates are obtained with infinite number of antennas $M$ and of reflecting elements $N$, when the RIS is aligned to a specific user and when the RIS is aligned to none of the users. Besides, we study the power scaling laws of the users. We find that while guaranteeing a required system performance, the transmit power of the users can be scaled down at most by the factor $\frac{1}{M}$ when $M$ goes infinite, or by the factor $\frac{1}{MN}$ when $M$ and $N$ go infinite.
  \item We propose an optimization algorithm based on GA, which can be applied to solving the continuous and the discrete phase shift optimization problems for maximizing the sum rates of the system. Moreover, we verify in simulations that the proposed optimization algorithm can largely improve the sum rates of the considered system.
\end{itemize}

The remainder of this paper is organized as follows:
Section II models the RIS-assisted multi-user uplink massive MIMO system under Rician fading channels and with imperfect hardware.
Section III derives the approximate expressions for ergodic AR, and analyzes the system performance.
Section IV proposes an algorithm based on GA to solve the continuous and discrete phase shift optimization problems for maximizing the sum rates of the system.
Section V provides the simulation results.
Section VI gives a brief conclusion.

\emph{Notations:}
In this paper, we use lower case letters, bold lower case letters and bold upper case letters to denote scalars, vectors and matrices, respectively.
The matrix inverse, conjugate-transpose, transpose and conjugate operations are respectively denoted by the superscripts ${\left(  \cdot  \right)^ {-1} }$, ${\left(  \cdot  \right)^H}$, ${\left(  \cdot  \right)^T}$ and ${\left(  \cdot  \right)^{*}}$.
We use ${\rm tr}\left(  \cdot  \right)$, $\left\|  \cdot  \right\|$ and ${\rm E}\left\{  \cdot  \right\}$ to denote trace, Euclidean 2-norm and the expectation operations, respectively.
And ${\left[ {\bf{A}} \right]_{ij}}$ denotes the $\left( {i,j} \right)$th element of matrix ${\bf{A}}$.
The matrix ${{\bf{I}}_N}$ denotes an $N \times N$ identity matrix.
In addition, we denote a circularly symmetric complex Gaussian vector ${\bf a}$ with zero mean and covariance ${\bf{\Sigma }}$ by ${\bf a} \thicksim {\cal C}{\cal N}\left( {{\bf{0}},{\bf{\Sigma }}} \right)$.

\section{System Model}

\begin{figure}
  \centering
  \includegraphics[scale=0.6]{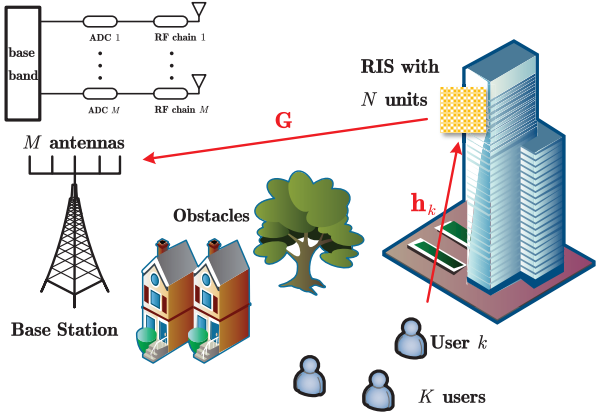}\\  
  \caption{System Model}\label{sys_model}  
\end{figure}

We investigate a multi-user uplink communication system with imperfect hardware, where $K$ single-antenna users transmit signals to a BS equipped with large-scale arrays of $M$ antennas.
As shown in Fig. \ref{sys_model}, the direct links between the users and the BS are blocked by obstacles such as buildings and trees. Therefore, we employ an RIS with $N$ elements to assist the communications between the users and the BS.

\subsection{Channel Model}

The channels from the users to the RIS and from the RIS to the BS follow the Rician fading distribution, which can be expressed respectively as
\begin{equation*} \vspace{-0.1cm}
  {\bf{H}} = \left[ {{{\bf{h}}_1},{{\bf{h}}_2},...,{{\bf{h}}_k}} \right],
\end{equation*}
\begin{equation}\label{channel_H} \vspace{-0.1cm}
     {\bf{h}}_k^{} = \sqrt {{\alpha _k}} \left( {\sqrt {\frac{{{\mu _k}}}{{{\mu _k} + 1}}} {\bf{\bar h}}_k^{} + \sqrt {\frac{1}{{{\mu _k} + 1}}} {\bf{\tilde h}}_k^{}} \right),
\end{equation}
\begin{equation}\label{channel_G} \vspace{-0.1cm}
  {\bf{G}} = \sqrt \beta  \left( {\sqrt {\frac{\delta }{{\delta  + 1}}} {\bf{\bar G}} + \sqrt {\frac{1}{{\delta  + 1}}} {\bf{\tilde G}}} \right),
\end{equation}
where scalars ${\alpha _k}$ and $\beta$ are respectively the large-scale fading coefficients between user $k$ and the RIS, and between the RIS and the BS.
Scalars ${\mu _k}$ and $\delta$ stand for the Rician factors of the channels between user $k$ and the RIS, and between the RIS and the BS, respectively.
Vector ${\bf{\tilde h}}_k  \in {\mathbb{C}^{N \times 1}}$ and matrix ${\bf{\tilde G}} \in {\mathbb{C}^{M \times N}}$ are the non-line-of-sight (nLoS) parts of the channels with independently and identically distributed (i.i.d) elements following the distribution of $\mathcal{CN}\left( 0,1 \right)$.
Vector ${\bf{\bar h}}_k  \in {\mathbb{C}^{N \times 1}}$ and matrix ${\bf{\bar G}} \in {\mathbb{C}^{M \times N}}$ are the LoS parts of the channels,
which are expressed respectively as \vspace{-0.05cm}
\begin{equation}\label{channel_h_LoS} \vspace{-0.05cm}
  {{{\bf{\bar h}}}_k} = {\bf{a}}_N^{}\left( {\phi _{kr}^a,\phi _{kr}^e} \right),
\end{equation}
\begin{equation}\label{channel_G_LoS} \vspace{-0.05cm}
  {\bf{\bar G}} = {{\bf{a }}_M}\left( {\phi _r^a,\phi _r^e} \right){\bf{a }}_N^H\left( {\phi _t^a,\phi _t^e} \right),
\end{equation}
where $\phi _{kr}^a$ and $\phi _{kr}^e$ represent the azimuth angles of arrival (AoA) and elevation AoA at the RIS from user $k$, respectively.
$\phi _t^a$ and $\phi _t^e$ are the azimuth angles of departure (AoD) and elevation AoD from the RIS to the BS, respectively.
$\phi _r^a$ and $\phi _r^e$ stand for the azimuth AoA and elevation AoA at the BS from the RIS, respectively.
Furthermore, it is assumed that uniform square planar arrays are equipped at both the RIS and the BS with size of $\sqrt{N} \times \sqrt{N}$ and $\sqrt{M} \times \sqrt{M}$, respectively.
Thus, the \textit{i}th element of vector ${\bf{a }}_X \in {\mathbb{C}^{N \times 1}}$ is expressed as \cite{9743440}, \cite{9079457}  \vspace{-0.05cm}
\begin{equation*} \vspace{-0.2cm}
  {\left[ {{\bf{a}}_X^{}\left( {{\phi _1},{\phi _2}} \right)} \right]_i} = {e^{j2\pi \frac{d}{\lambda }\left( {{x_i}\sin {\phi _1}\sin {\phi _2} + {y_i}\cos {\phi _2}} \right)}},
\end{equation*}
\begin{equation}\label{array_aXi}  \vspace{-0.05cm}
     {x_i} = \left( i-1 \right) \bmod \sqrt X , \;\;  {y_i} = \left\lfloor {\frac{i-1}{{\sqrt X }}} \right\rfloor ,
\end{equation}
where $d$ is the antenna/unit spacing, and $\lambda$ is the carrier wavelength.

\begin{figure*}[b]
\hrulefill 
\setcounter{equation}{13}
\begin{equation}\label{EAR_user_k}
  R_k=\mathrm{E}\left\{ \log _2\left( 1+\frac{p_k\tau ^2\left| \mathbf{h}_{k}^{H}\mathbf{\Phi }^H\mathbf{G}^H \bm{\chi} \mathbf{G\Phi \Theta h}_k \right|^2}{\sum_{i\ne k}^K{p_i\tau ^2\left| \mathbf{h}_{k}^{H}\mathbf{\Phi }^H\mathbf{G}^H \bm{\chi} \mathbf{G\Phi \Theta h}_i \right|^2}+\mathrm{DN}_k+\mathrm{AN}_k+\mathrm{QN}_k} \right) \right\}
\end{equation}
\setcounter{equation}{5}

\hrulefill 
\setcounter{equation}{17}
\begin{equation}\label{EAR_user_k_approx}
  \tilde{R}_k=\log _2\left( 1+\frac{p_k\tau ^2\xi _k}{\sum_{i\ne k}^K{p_i\tau ^2\gamma _{ki}}+\tau \left( 1-\tau \right) \varpi _k\zeta M+\tau \left( \sigma _{\mathrm{RF}}^{2}+\sigma ^2 \right) \varpi _kM} \right)
\end{equation}
\setcounter{equation}{5}

\hrulefill 
\setcounter{equation}{18}
\begin{equation}\label{rayleigh_EAR_user_k_approx}
  \tilde{R}_k=\log _2\left( 1+\frac{\left( \rho ^2N+2-\rho ^2 \right) \iota ^2M+\left( \left( 1-\iota ^2 \right) \rho ^2+1 \right) N+\left( \iota ^2-1 \right) \rho ^2+3-6\iota ^2}{+\sum_{i=1}^K{\frac{p_i\alpha _i}{p_k\alpha _k}\left( \iota ^2M+1-\iota ^2+\frac{1-\tau}{\tau}N \right)}+\frac{\sigma _{\mathrm{RF}}^{2}+\sigma ^2}{p_k\tau \kappa ^2\beta \alpha _k}-\iota ^2M-1+\iota ^2} \right)
\end{equation}
\setcounter{equation}{5}
\end{figure*}

%
%

\subsection{Data Transmission with Imperfect Hardware}

Since the RIS is employed to assist the communications between the users and the BS, the signals are first transmitted to the RIS, then reflected to the BS.
Therefore, when the RIS and the BS have ideal configurations, the received signal at the BS can be expressed as \vspace{-0.05cm}
\begin{equation}\label{y_noHI} \vspace{-0.05cm}
  {\bf{y}}_{\rm p} = {\bf{G\Phi HPx}} + {\bf{n}},
\end{equation}
where ${\bf{x}} = \left[ {{x_1},{x_2},...,{x_K}} \right]_K^T$ with $x_k$ representing the signal transmitted from user $k$ and subject to ${\rm E}\left\{ {{{\left| {{x_k}} \right|}^2}} \right\} = 1$.
${\bf{P}} = {\rm{diag}}\left\{ {{p_1},{p_2},...,{p_K}} \right\}^{{1 \mathord{\left/
 {\vphantom {1 2}} \right.
 \kern-\nulldelimiterspace} 2}}$, and $p_k$ is the transmit power of user $k$.
${\bf{\Phi }} = {\rm{diag}}{\left\{ {{e^{j{\theta _1}}},{e^{j{\theta _2}}},...,{e^{j{\theta _N}}}} \right\}}$ stands for the phase shift matrix, and $\theta _n$ is the phase shift at the unit $n$ of the RIS.
${\bf{n}}$ is the additive white Gaussian noise (AWGN) at the BS, whose elements follow i.i.d $\mathcal{CN}\left( 0,\sigma^2 \right)$.

However, in real scenarios, the imperfect hardware at the RIS and the BS should be taken into consideration , which would limit the system performance.
To address this, we consider the phase noise at the RIS, the RF impairments and low-resolution ADCs at the BS.

The phase noise is induced by the imperfection of the reflecting elements, or by the imperfect channel estimation \cite{8869792}.
In this paper, it is modelled as
\begin{equation}\label{phase_noiise_model}
  {\bf \Theta}  = {\rm{diag}}{\left\{ {{e^{j{\varepsilon _1}}},{e^{j{\varepsilon _2}}},...,{e^{j{\varepsilon _N}}}} \right\}},
\end{equation}
where ${\varepsilon _n}$ follows a zero-mean Von Mises distribution with the probability density function (PDF) of \cite{9534477} \vspace{-0.05cm}
\begin{equation}\label{pdf_VMs_vn} \vspace{-0.05cm}
  M_{\upsilon}\left( \varepsilon_n \right) =\frac{1}{2\pi I_0\left( \upsilon \right)}e^{\upsilon \cos \varepsilon_n},\;\;\; \varepsilon_n \in \left[ 0,2\pi \right),
\end{equation}
with $\upsilon$ being the concentration parameter.

Then, we adopt the extended error vector magnitude (EEVM) model to depict the impacts of the RF impairments, such as I/Q imbalance and carrier frequency offset \cite{4537203,8629287,9079457}.
Thus, the received signal at the RF chains is re-expressed as
\begin{equation}\label{y_RF}
  \mathbf{y}_{\mathrm{RF}}= \bm{\chi} \mathbf{ G\Phi \Theta HPx}+\mathbf{n}_{\mathrm{RF}}+\mathbf{n},
\end{equation}
where $\bm{\chi }=\mathrm{diag}\left\{ \chi _1,\chi _2,...,\chi _M \right\}$ with $\chi _m=\kappa _me^{j\varphi _m}$ representing the amplitude attenuation and phase shift for the $m$th RF chain. The scalar $\kappa _m$ satisfies $\left| \kappa _m \right|\leqslant 1$, while $\varphi _m$ follows a uniform distribution of $\mathcal{U} \left[ -\eta _m,\eta _m \right]$ with $\eta _m\in \left[ 0,\pi \right)$.
Additionally, vector $\mathbf{n}_{\mathrm{RF}}=\left[ n_{\mathrm{RF},1},n_{\mathrm{RF},2},...,n_{\mathrm{RF},M} \right] ^T$ stands for the additive distortion noise at the RF chains, where $n_{\mathrm{RF},m}$ has the distribution of $\mathcal{C} \mathcal{N} \left( 0,\sigma _{\mathrm{RF},m}^{2} \right)$.
In the following analysis, we assume $\kappa _m=\kappa$, $\eta _m=\eta$ and $\sigma _{\mathrm{RF},m}^{2}=\sigma _{\mathrm{RF}}^{2}$ for $\forall m$ for simplicity, which means all the RF chains have the same level of imperfections.

Moreover, the BS uses low-resolution ADCs to reduce the hardware cost and power consumption, the impacts of which can be modelled by the additive quantization
noise model (AQNM) \cite{7979627,8629287}. Therefore, the quantized signals can be obtained as
\begin{equation}\label{y_Q}
  \mathbf{y}_{\mathrm{Q}}=\tau \bm{\chi} \mathbf{ G\Phi \Theta HPx}+\tau \mathbf{n}_{\mathrm{RF}}+\tau \mathbf{n}+\mathbf{n}_{\mathrm{Q}},
\end{equation}
where $\tau =1-\varrho$, and $\varrho$ is the inverse of the signal-to-quantization-noise ratio.
For the quantization bits $b \leq 5$, the values of $\varrho$ are listed in Table \ref{values_for_varrho}, while for $b>5$, we have $\varrho =\frac{\pi \sqrt{3}}{2}\cdot 2^{-2b}$.
Vector $\mathbf{n}_{\mathrm{Q}}\sim \mathcal{C} \mathcal{N} \left( \mathbf{0},\tau \left( 1-\tau \right) \mathbf{S}_{\mathrm{Q}} \right)$ represents the additive Gaussian quantization noise, which is uncorrelated with $\mathbf{y}_{\mathrm{RF}}$.
The matrix $\mathbf{S}_{\mathrm{Q}}$ satisfies $\mathbf{S}_{\mathrm{Q}}=\mathrm{diag}\left\{ \mathrm{E}\left\{ \mathbf{y}_{\mathrm{RF}}\mathbf{y}_{\mathrm{RF}}^{H} \right\} \right\}$, which yields
\begin{equation}\label{S_Q_mm}
  \left[ \mathbf{S}_{\mathrm{Q}} \right] _{mm}=\mathrm{E}\left\{ \left| \mathbf{1}_{M,m}\bm{\chi}\mathbf{G\Phi \Theta HPx} \right|^2 \right\} +\sigma _{\mathrm{RF}}^{2}+\sigma ^2,
\end{equation}
where ${\bf{1}}_{M,m} \in \mathbb{C}^{1 \times M}$ is the vector whose \textit{m}th element is 1, while the rest elements are zero.

\begin{table}
\caption{Values of $\varrho$}\label{values_for_varrho} \vspace{0.1cm}
\renewcommand\arraystretch{1.6}
\begin{tabular}{|c|c|c|c|c|c|}
\hline
 $b$    & 1      & 2      & 3       & 4        & 5        \\ \hline
 $\varrho$    & 0.3634 & 0.1175 & 0.03454 & 0.009497 & 0.002499 \\ \hline
\end{tabular}
\end{table}

After the quantization, the BS adopts the maximal-ratio-combining (MRC) processing. Then, the processed signal is obtained as \vspace{-0.05cm}
\begin{equation}\label{Wy_HI} \vspace{-0.05cm}
  \mathbf{r} \!=\! \mathbf{Wy}_{\mathrm{Q}} \!=\! \tau \mathbf{W} \bm{\chi} \mathbf{G\Phi \Theta HPx}+\tau \mathbf{Wn}_{\mathrm{RF}}+\tau \mathbf{Wn}+\mathbf{Wn}_{\mathrm{Q}},
\end{equation}
where the beamforming matrix is given by ${\bf{W}} = {{\bf{H}}^H}{{\bf{\Phi }}^H}{{\bf{G}}^H}$.
Herein, we focus on the signal transmitted from user $k$, which is obtained as \vspace{0.05cm}
\begin{equation*} \vspace{-0.05cm}
  \hspace{-3.5cm} r_k=\sqrt{p_k}\tau \mathbf{h}_{k}^{H}\mathbf{\Phi }^H\mathbf{G}^H\bm{\chi} \mathbf{G\Phi \Theta h}_kx_k
\end{equation*}
\begin{equation*}
  +\sum_{i\ne k}^K{\sqrt{p_i}\tau \mathbf{h}_{k}^{H}\mathbf{\Phi }^H\mathbf{G}^H\bm{\chi} \mathbf{G\Phi \Theta h}_ix_i}+\tau \mathbf{h}_{k}^{H}\mathbf{\Phi }^H\mathbf{G}^H\mathbf{n}_{\mathrm{RF}}
\end{equation*}
\begin{equation}\label{Wy_HI_user_k}
  \hspace{-3.1cm} +\tau \mathbf{h}_{k}^{H}\mathbf{\Phi }^H\mathbf{G}^H\mathbf{n}+\mathbf{h}_{k}^{H}\mathbf{\Phi }^H\mathbf{G}^H\mathbf{n}_{\mathrm{Q}}.
\end{equation}
Note that at the right hand side of \eqref{Wy_HI_user_k}, the first term is the desired signal from user $k$;
the second term is the multi-user interference from the other users;
the third and the fourth terms are generated from the distortion noise and the AWGN at RF chains, respectively;
the last term is from the quantization noise.

\section{Achievable Rate Analysis}
In this section, we investigate the uplink ergodic AR of this multi-user massive MIMO system and the impacts of the system settings, such as the number of the reflecting elements $N$ at the RIS and the number of the antennas $M$ at the BS.

\addtocounter{equation}{1}
According to \eqref{Wy_HI_user_k}, the uplink ergodic AR for user $k$ is expressed as \eqref{EAR_user_k} at the bottom of this page,
where the power of the distortion noise, the AWGN, and the quantization noise are respectively expressed as \vspace{-0.05cm}
\begin{equation}\label{DN_k} \vspace{-0.05cm}
  \mathrm{DN}_k=\tau ^2\sigma _{\mathrm{RF}}^{2}\left\| \mathbf{h}_{k}^{H}\mathbf{\Phi }^H\mathbf{G}^H \right\| ^2,
\end{equation}
\begin{equation}\label{AN_k} \vspace{-0.05cm}
  \mathrm{AN}_k=\tau ^2\sigma ^2\left\| \mathbf{h}_{k}^{H}\mathbf{\Phi }^H\mathbf{G}^H \right\| ^2,
\end{equation}
\begin{equation}\label{QN_k} \vspace{-0.05cm}
  \mathrm{QN}_k=\tau \left( 1-\tau \right) \left\| \mathbf{h}_{k}^{H}\mathbf{\Phi }^H\mathbf{G}^H\mathbf{S}_{\mathrm{Q}}^{{{1}/{2}}} \right\| ^2.
\end{equation}
Based on \eqref{EAR_user_k}, we present the following theorem.

\begin{theorem} \label{theorem_Rk_approx}
\addtocounter{equation}{1}
With MRC processing, the uplink ergodic AR for user $k$ in the RIS-assisted multi-user uplink massive MIMO system can be approximated in a closed form as \eqref{EAR_user_k_approx} at the bottom of this page,
where ${\xi _k}$, ${\gamma _{ki}}$, ${\varpi _k}$ and $\zeta$ are respectively given by \eqref{hk_hk_norm2}, \eqref{hk_hi_norm2}, \eqref{bar_w_k} and \eqref{zeta_value} in Appendix \ref{derivation_Rk_approx}.
\end{theorem}

\begin{IEEEproof}
See Appendix \ref{derivation_Rk_approx}.
\end{IEEEproof}

It can be readily observed from \emph{Theorem} \ref{theorem_Rk_approx} that the uplink ergodic AR of user $k$  depends on $M$, $N$, the AoA at the BS, the AoA and AoD at the RIS, the power of users, the large-scale fading coefficients, the Rician factors, the phase shifts and noise at the RIS, the RF impairments, and the quantization bit and accuracy.

In particular, when the Rician factors satisfy $\mu _k=0$, $\forall k$ and $\delta =0$, the Rician fading channels expressed in \eqref{channel_H} and \eqref{channel_G} degenerate to Rayleigh fading channels, where only the nLoS parts of the channels remain. The following remark discusses the uplink ergodic AR with Rayleigh fading channels under the considered system.

\begin{remark}\label{cor_rayleigh}
\addtocounter{equation}{1}
When Rayleigh fading channels are considered, where $\mu _k=0$, $\forall k$ and $\delta =0$, the closed-form expression of the uplink ergodic AR for user $k$ in the RIS-assisted multi-user uplink massive MIMO system is obtained as \eqref{rayleigh_EAR_user_k_approx} at the bottom of the previous page, where $\rho$ and $\iota$ are respectively given by \eqref{rho_value} and \eqref{iota_definition} in Appendix \ref{derivation_Rk_approx}.
\end{remark}


\subsection{Asymptotic Analysis}

According to \eqref{fk} in Appendix \ref{derivation_Rk_approx}, it is noted that $\left| f_k \right|\leqslant N$ and the equality holds when the RIS is aligned to user $k$. Specifically, the phase shifts aligned to user $k$ at the RIS should satisfy
\begin{equation*}
  \theta _n=-2\pi \frac{d}{\lambda}\left( x_np_k+y_nq_k \right) ,\;\;\;  \forall n ,
\end{equation*}
\begin{equation*}
  p_k=\sin \phi _{kr}^{a}\sin \phi _{kr}^{e}-\sin \phi _{t}^{a}\sin \phi _{t}^{e} ,
\end{equation*}
\begin{equation}\label{aligned_k_constraints}
  q_k=\cos \phi _{kr}^{e}-\cos \phi _{t}^{e} .
\end{equation}
In this case, we assume that $\left| f_i \right|$ is bounded when $i \neq k$.
Therefore, when the RIS is aligned to user $k$, the numerator ${p_k}{\xi _k}$ in \eqref{EAR_user_k_approx} is on the order of $\mathcal{O} \left( M^2N^4 \right)$.

To obtain the order of the denominator in \eqref{EAR_user_k_approx}, we need to focus on the term $\mathbf{\bar{h}}_{k}^{H}\mathbf{\bar{h}}_i$ in \eqref{zki_1} and \eqref{zki_4} first, which can be further expressed as
\begin{equation*} \vspace{-0.2cm}
  \hspace{-2.5cm} \mathbf{\bar{h}}_{k}^{H}\mathbf{\bar{h}}_i=\sum_{n=1}^N{a_{Nn}^{\ast}\left( \phi _{kr}^{a},\phi _{kr}^{e} \right) a_{Nn}^{}\left( \phi _{ir}^{a},\phi _{ir}^{e} \right)}
\end{equation*}
\begin{equation*}  \vspace{-0.2cm}
  =\sum_{n=1}^N{e^{j2\pi \frac{d}{\lambda}\left( x_n\left( \sin \phi _{ir}^{a}\sin \phi _{ir}^{e} \!-\sin \phi _{kr}^{a}\sin \phi _{kr}^{e} \right) +y_n\left( \cos \phi _{ir}^{e} \!-\cos \phi _{kr}^{e} \right) \right)}}
\end{equation*}
\begin{equation*} \vspace{-0.2cm}
  \hspace{-0.4cm} \triangleq \sum_{n=1}^N{e^{j2\pi \frac{d}{\lambda}\left( x_ns_{ik}+y_nt_{ik} \right)}} \overset{\left( a \right)}{=} \sum_{y=0}^{\sqrt{N}-1}{\sum_{x=0}^{\sqrt{N}-1}{e^{j2\pi \frac{d}{\lambda}\left( xs_{ik}+yt_{ik} \right)}}}
\end{equation*}
\begin{equation}\label{expand_hbark_hbari} \vspace{-0.05cm}
  \hspace{-3.2cm} =\sum_{y=0}^{\sqrt{N}-1}{e^{j2\pi \frac{d}{\lambda}\left( yt_{ik} \right)}}\sum_{x=0}^{\sqrt{N}-1}{e^{j2\pi \frac{d}{\lambda}\left( xs_{ik} \right)}}.
\end{equation}
Step $\left( a \right)$ is because $x_n=\left( n-1 \right) \mathrm{mod}\sqrt{N}, \left. y_n=\left. \lfloor \frac{n-1}{\sqrt{N}} \right. \right. \rfloor$. On the other hand, for fixed $s$, we have
\begin{equation*}
  \sum_{x=0}^{\sqrt{N}-1} \!\!\! {e^{j2\pi \frac{d}{\lambda}xs}} \!\!=\!\! \frac{1 \!\!-\! e^{j2\pi \frac{d}{\lambda}\sqrt{N}s}}{1 \!\!-\! e^{j2\pi \frac{d}{\lambda}s}} \!\!=\!\! \frac{(\! e^{-j\pi \frac{d}{\lambda}\sqrt{N}s} \!\!-\! e^{j\pi \frac{d}{\lambda}\sqrt{N}s} \!) e^{j\pi \frac{d}{\lambda}\sqrt{N}s}}{(\! e^{-j\pi \frac{d}{\lambda}s} \!\!-\! e^{j\pi \frac{d}{\lambda}s} \!) e^{j\pi \frac{d}{\lambda}s}}
\end{equation*}
\begin{equation}\label{}
  =\!\! \frac{\sin\mathrm{(}\pi \frac{d}{\lambda}\sqrt{N}s)}{\sin\mathrm{(}\pi \frac{d}{\lambda}s)}e^{j\pi \frac{d}{\lambda}(\sqrt{N} -\! 1)s} \!\!=\!\! \sqrt{N}\frac{\sin\mathrm{c(}\frac{d}{\lambda}\sqrt{N}s)}{\sin\mathrm{c(}\frac{d}{\lambda}s)}e^{j\pi \frac{d}{\lambda}(\sqrt{N} -\! 1)s}
\end{equation}
which yields
\begin{equation}\label{}
\frac{\sum_{x=0}^{\sqrt{N}-1}{e^{j2\pi \frac{d}{\lambda}xs}}}{\sqrt{N}}\xrightarrow{N\rightarrow \infty}0.
\end{equation}
According to \eqref{expand_hbark_hbari}, it is noted that $\mathbf{\bar{h}}_{k}^{H}\mathbf{\bar{h}}_i$ is below the order of $\mathcal{O} \left( N \right)$, which means $\left| \mathbf{\bar{h}}_{k}^{H}\mathbf{\bar{h}}_i \right|^2$ is not the term which dominates the order of the denominator.
Since the quantization noise is proportional to the power of the received signals, the denominator is on the order of $\mathcal{O} \left( MN^4 \right)$.
Thus, the fraction in \eqref{EAR_user_k_approx} is on the order of $\mathcal{O} \left( M \right)$.
Then, we have the following corollary:

\begin{corollary}\label{cor_Rk_N_infinity}
When the RIS is aligned to user $k$ and the number of the reflecting elements at the RIS satisfies $N \rightarrow \infty$, the uplink ergodic AR for user $k$ in the RIS-assisted multi-user uplink massive MIMO system converges to
\begin{equation}\label{EAR_user_k_approx_N_infinity}
  \tilde{R}_k\rightarrow \log _2\left( \frac{\tau \iota ^2}{1-\tau}M+\frac{1-\iota ^2\tau}{1-\tau} \right),
\end{equation}
where $\iota$ is given by \eqref{iota_definition} in Appendix \ref{derivation_Rk_approx}.
\end{corollary}

It can be seen from \emph{Corollary} \ref{cor_Rk_N_infinity} that the converged uplink ergodic AR for user $k$ is determined by $M$, the quantization parameter $\tau$, and the scalar $\iota$ which is related to the RF impairments.

Furthermore, we can find that the impact of the phase noise at the RIS vanishes in the converged AR for user $k$. This is because when the RIS is aligned to user $k$, the signal transmitted from user $k$ is on the order of $\mathcal{O} \left( N^4 \right)$, while the interference from the other users is on the order of $\mathcal{O} \left( N^3 \right)$. Then as $N$ goes to infinity, the signal transmitted from user $k$ becomes dominant. On the other hand, the phase noise at the RIS is multiplicative to the transmitted signal, and the quantization noise at the ADCs is proportional to the signal power. Therefore, the impact of phase noise at the RIS vanishes as the equal parts of coefficients in the numerator and denominator are eliminated.


When none of the users is aligned to the RIS, we assume that $\left| f_k \right|$ is bounded for $\forall k$.
In this case, both of the numerator and the denominator in the fraction of \eqref{EAR_user_k_approx} are on the order of $\mathcal{O} \left( M^2N^2 \right)$.
Then, we have the following conclusion.

\begin{corollary}\label{cor_MN_infinity}
When none of the users is aligned to the RIS, as the number of the reflecting elements $N$ at the RIS and the number of the antennas $M$ at the BS satisfy $N,M \rightarrow \infty$, the uplink ergodic AR for user $k$ in the RIS-assisted multi-user uplink massive MIMO system converges to
\begin{equation}\label{MN_infinity}
  \tilde{R}_k \! \rightarrow \log _2 \! \left( \! 1  \!+\! \frac{\frac{\rho ^2}{\delta ^2}\left( \mu _k+\delta +1 \right) ^2+\mu _k+1-\rho ^2\mu _k}{\sum_{i\ne k}^K{\frac{p_i\alpha _i\left( \mu _k+1 \right)}{p_k\alpha _k\left( \mu _i+1 \right)}\left( \mu _i+1-\rho ^2\mu _i \right)}} \right)  ,
\end{equation}
where $\rho$ is given by \eqref{rho_value} in Appendix \ref{derivation_Rk_approx}.
\end{corollary}

In \emph{Corollary} \ref{cor_MN_infinity}, the converged uplink ergodic AR for user $k$ is determined by the power of the users, the large-scale fading coefficients, the Rician factors, and the level of phase noise.

\begin{figure*}[b]
\hrulefill 
\setcounter{equation}{29}
\begin{equation}\label{Rk_scaling_MN_none_law}
  \tilde{R}_k\rightarrow \log _2\left( 1+\frac{\frac{\rho ^2}{\delta ^2}\left( \mu _k+\delta +1 \right) ^2+\mu _k+1-\rho ^2\mu _k}{\sum_{i\ne k}^K{\frac{\alpha _i\left( \mu _k+1 \right)}{\alpha _k\left( \mu _i+1 \right)}\left( \mu _i+1-\rho ^2\mu _i \right)}+\frac{\left( \sigma _{\mathrm{RF}}^{2}+\sigma ^2 \right) \left( \mu _k+\delta +1 \right) \left( \delta +1 \right) \left( \mu _k+1 \right)}{E_u\iota ^2\kappa ^2\tau \beta \alpha _k\delta ^2}} \right)
\end{equation}
\setcounter{equation}{25}

\hrulefill 
\setcounter{equation}{30}
\begin{equation}\label{Rk_scaling_MN_k_law}
  \tilde{R}_k\rightarrow \log _2\left( 1+\frac{\rho ^2\mu _k}{\sum_{i\ne k}^K{\frac{\alpha _i\left( \mu _k+1 \right)}{\alpha _k\left( \mu _i+1 \right)}\left( \mu _i+1-\rho ^2\mu _i \right)}+\left( \sigma _{\mathrm{RF}}^{2}+\sigma ^2 \right) \frac{\left( \delta +1 \right) \left( \mu _k+1 \right)}{E_u\tau \iota ^2\kappa ^2\beta \alpha _k\delta}} \right)
\end{equation}
\setcounter{equation}{25}
\end{figure*}

\subsection{Power Scaling Laws}
An important feature of massive MIMO is that it reduces the transmit power of the users proportionally to the number of antennas while maintaining the required system performance. As such, to gather valuable insights on energy savings, we investigate the power scaling laws of the users..

We scale down the power as $p_k=\frac{E_u}{M^{\epsilon}}$, $\forall k$, with fixed $E_u$, where $\epsilon \geqslant 0$ is the power scaling factor deciding the power scaling level.
Then, from \eqref{EAR_user_k_approx} in \emph{Theorem} \ref{theorem_Rk_approx}, as the number of antennas $M\rightarrow \infty$, we have
\begin{equation}\label{Rk_scaling_M}
\tilde{R}_k\xrightarrow{M\rightarrow \infty}\begin{cases}
	0,\epsilon >1\\
	\log _2\left( 1+\frac{E_u\varGamma _k}{\sum_{i\ne k}^K{E_u\varGamma _{ki}}+\left( \sigma _{\mathrm{RF}}^{2}+\sigma ^2 \right) \tau \varpi _k} \right) ,\epsilon =1\\
	\log _2\left( 1+\frac{\varGamma _k}{\sum_{i\ne k}^K{\varGamma _{ki}}} \right) ,\epsilon <1\\
\end{cases}
\end{equation}
where $\varGamma _k$ and $\varGamma _{ki}$ are respectively defined as \eqref{varGamma_k} and \eqref{varGamma_ki}
\begin{equation*} \vspace{-0.2cm}
  \hspace{-4.8cm} \varGamma _k =  \frac{\tau ^2\iota ^2\kappa ^2\beta ^2\alpha _{k}^{2}}{\left( \delta +1 \right) ^2\left( \mu _k+1 \right) ^2} \times
\end{equation*}
\begin{equation*} \vspace{-0.2cm}
  \hspace{-1.8cm} \Bigg( \left( \rho ^2\left( \mu _k+\delta +1 \right) ^2+\left( 1-\rho ^2 \right) \delta ^2\mu _k+\delta ^2 \right) N^2
\end{equation*}
\begin{equation*} \vspace{-0.2cm}
  \hspace{-0.8cm} + \Big( \left( \left( 2\mu _k+3\delta +2-\delta \mu _k \right) \rho ^2+\left( 1+\mu _k \right) \delta \right) \delta \mu _k\left| f_k \right|^2
\end{equation*}
\begin{equation*} \vspace{-0.2cm}
  + \left( \mu _k+\delta +2 \right) ^2-\rho ^2\left( \mu _k+\delta +1 \right) ^2-2\rho ^2\delta \mu _k-2 \Big) N
\end{equation*}
\begin{equation}\label{varGamma_k} \vspace{-0.1cm}
  + \rho ^2\delta ^2\mu _{k}^{2}\left| f_k \right|^4+2\left( \left( 1-\rho ^2 \right) \left( \mu _k+\delta \right) +2 \right) \delta \mu _k\left| f_k \right|^2 \Bigg) ,
\end{equation}
\begin{equation*} \vspace{-0.2cm}
  \varGamma _{ki} =  \frac{\tau ^2\iota ^2\kappa ^2\beta ^2\alpha _k\alpha _i}{\left( \delta \!+\! 1 \right) ^2\left( \mu _k \!+\! 1 \right) \left( \mu _i \!+\! 1 \right)} \times
  \Bigg( \left( \mu _i+1-\rho ^2\mu _i \right) \delta ^2N^2
\end{equation*}
\begin{equation*} \vspace{-0.2cm}
  \hspace{-1.8cm} + \Big( \left( \mu _i+1-\rho ^2\mu _i \right) \delta ^2\mu _k\left| f_k \right|^2+\rho ^2\delta ^2\mu _i\left| f_i \right|^2
\end{equation*}
\begin{equation*} \vspace{-0.2cm}
  + \left( \mu _k+2\delta +1 \right) \left( \mu _i+1-\rho ^2\mu _i \right) +\rho ^2\mu _i \Big) N
\end{equation*}
\begin{equation*} \vspace{-0.2cm}
  \hspace{-2cm} +\left( \rho ^2\delta \mu _i\left| f_i \right|^2+2\mu _i\left( 1-\rho ^2 \right) +2 \right) \delta \mu _k\left| f_k \right|^2
\end{equation*}
\begin{equation}\label{varGamma_ki}
  \hspace{-0.3cm} +\left( 2\delta \left| f_i \right|^2 \!+\! \mu _k\left| \mathbf{\bar{h}}_{k}^{H}\mathbf{\bar{h}}_i \right|^2 \!+\! 2\delta \mu _k\mathrm{Re}\left( f_{k}^{\ast}f_i\mathbf{\bar{h}}_{i}^{H}\mathbf{\bar{h}}_k \right) \right) \rho ^2\mu _i \Bigg),
\end{equation}
and $\rho$, $\iota$, and $\varpi _k$ are respectively given by \eqref{rho_value}, \eqref{iota_definition}, and \eqref{bar_w_k} in Appendix \ref{derivation_Rk_approx}.

It is noted that the converged AR is determined by the power scaling factor $\epsilon$.
When $\epsilon > 1$, the converged AR goes to zero as $M\rightarrow \infty$. This is because the transmit power is aggressively scaled down with a large $\epsilon$, which certainly degrades the system performance.
When $\epsilon \leqslant 1$, the converged AR is a non-zero value as $M\rightarrow \infty$. It means the system performance can be maintained when the transmit power is scaled down.
Furthermore, when $\epsilon = 1$, the transmit power can be scaled down by the factor $\frac{1}{M}$ at most while guaranteeing the required system performance.
Therefore, we obtain the following corollary.

\begin{corollary}\label{scaling_M}
As the number of antennas $M\rightarrow \infty$, the transmit power of the users can be scaled down at most to $p_k=\frac{E_u}{M}$, $\forall k$, with fixed $E_u$, then the uplink ergodic AR for user $k$ in the RIS-assisted multi-user uplink massive MIMO system converges to
\begin{equation}\label{Rk_scaling_M_law}
  \tilde{R}_k\rightarrow \log _2\left( 1+\frac{E_u\varGamma _k}{\sum_{i\ne k}^K{E_u\varGamma _{ki}}+\left( \sigma _{\mathrm{RF}}^{2}+\sigma ^2 \right) \tau \varpi _k} \right),
\end{equation}
where $\varGamma _k$ and $\varGamma _{ki}$ are respectively defined in \eqref{varGamma_k} and \eqref{varGamma_ki}.
\end{corollary}

\emph{Corollary} \ref{scaling_M} illustrates the power scaling law related to $M$. Theoretically, the transmit power of the users can be further cut down according to the number of reflecting elements $N$ at the RIS in the considered RIS-assisted system. The following corollary shows the power scaling law related to $M$ and $N$, when none of the users is aligned to the RIS.

\begin{corollary}\label{scaling_MN_none}
\addtocounter{equation}{1}
When none of the users is aligned to the RIS, as the number of antennas and reflecting elements $M,N\rightarrow \infty$, the transmit power of the users can be scaled down at most to $p_k=\frac{E_u}{MN}$, $\forall k$, with fixed $E_u$, then the uplink ergodic AR for user $k$ in the RIS-assisted multi-user uplink massive MIMO system converges to \eqref{Rk_scaling_MN_none_law} at the bottom of this page.
\end{corollary}

\emph{Corollary} \ref{scaling_MN_none} investigates the power scaling law in the case where none of the users is aligned to the RIS. Moreover, the following \emph{Corollary} \ref{scaling_MN_k} investigates the power scaling law in the case where the RIS is aligned to user $k$.

\begin{corollary}\label{scaling_MN_k}
\addtocounter{equation}{1}
When the RIS is aligned to user $k$, as the number of antennas and reflecting elements $M,N\rightarrow \infty$, the transmit power of user $k$ can be scaled down at most to $p_k=\frac{E_u}{MN^2}$, while the transmit power of other users is scaled down to $p_i=\frac{E_u}{MN}$, $i \neq k$, with fixed $E_u$, then the uplink ergodic AR for user $k$ in the RIS-assisted multi-user uplink massive MIMO system converges to \eqref{Rk_scaling_MN_k_law} at the bottom of this page.
\end{corollary}



\section{Phase Shift Optimization}
\emph{Theorem} \ref{theorem_Rk_approx} shows that the uplink ergodic AR depends on the phase shift of the RIS.
Thus, to enhance the system performance, we can optimize the phase shifts at the RIS for maximizing the sum uplink ergodic AR.

From \eqref{EAR_user_k_approx}, the sum uplink ergodic AR of the RIS-assisted multi-user uplink massive MIMO system can be expressed as \vspace{-0.05cm}
\begin{equation}\label{R_sum} \vspace{-0.05cm}
  \tilde{R}_{\mathrm{sum}}=\sum_{k=1}^K{\tilde{R}_k}.
\end{equation}
Since the phase shift at each unit of the RIS lies in the range of $\left[ 0,2\pi \right)$, the phase shift optimization problem can be expressed as
\begin{align}\label{phase_opt_continuous}
  \max_{\mathbf{\Phi }} \;\;\;  & \tilde{R}_{\mathrm{sum}}   \nonumber \\
  \mathrm{s}.\mathrm{t}.\;\;\;  & \theta _n\in \left[ 0,2\pi \right),\; \forall n ,
\end{align}
where $\tilde{R}_{\mathrm{sum}}$ is the sum uplink ergodic AR defined in \eqref{R_sum}, $\mathbf{\Phi }$ is the phase shift matrix, and $\theta _n$ is the phase shift at unit $n$ of the RIS.
According to \eqref{EAR_user_k_approx} and \eqref{R_sum}, the expression for the objective function of Problem \eqref{phase_opt_continuous} is complicated, which makes the problem challenging to be solved.
Herein, we apply GA to solve our phase shift optimization problem.

Simulating the evolution of nature population, GA mainly includes five parts: population initialization, fitness evaluation \& sort, selection, crossover and mutation.
By tailoring it to our phase shift optimization problem, we further discuss the five parts of GA.

\emph{1) Population initialization:}
In the beginning, $N_{\mathrm{tot}}$ individuals are generated in the initial population.
For the $N$ phase shifts at the $N$ reflecting elements of the RIS, each individual has one chromosome containing $N$ genes.
Furthermore, the value of each gene is initially generated in $\left[ 0,2\pi \right)$.

\emph{2) Fitness evaluation \& sort:}
The fitness value of the individual $i$ is evaluated by the fitness function, which is given by
\begin{equation}\label{fitness_function}
  f_{\mathrm{fit},i}=\tilde{R}_{\mathrm{sum},i}, \;\;\;  i=1,2,...,N_{\mathrm{tot}},
\end{equation}
where $\tilde{R}_{\mathrm{sum},i}$ is the sum uplink ergodic AR corresponding to individual $i$.
Since we use the objective function in Problem \eqref{phase_opt_continuous} as the fitness function, the individual corresponding to a higher sum uplink ergodic AR has a higher fitness value.
Then, $N_{\mathrm{tot}}$ individuals are sorted by the fitness values given in \eqref{fitness_function} as LIST 1.

\begin{algorithm}   
\caption{Algorithm for selection} 
\label{selection_ALG} 
\begin{algorithmic}[1] 
\State Generate $c$ from a uniform distribution in $\left( 0,1 \right)$;

\State Set $\mathrm{acc}=\bar{f}_{\mathrm{fit},j}$;

\For{$j=1:N_{\mathrm{tot}}$}
    \If{$c>\mathrm{acc}$}
       \State $j=j+1$;
       \State $\mathrm{acc}=\mathrm{acc}+\bar{f}_{\mathrm{fit},j}$;
    \Else
       \State Set the $j$th individual in the LIST 1 as a parent;
       \State break;
    \EndIf
\EndFor

\end{algorithmic}
\end{algorithm}

\emph{3) Selection:}
The selection operation is to choose individuals from the current population as parents.
Before the selection operation, we choose the top $N_{\mathrm{e}}$ individuals in LIST 1 as elite individuals, which are directly copied to the next generation. The operation for the elite individuals is to make sure that the optimal individuals in the current population are passed to the next.
Then, we scale the fitness values of the individuals in LIST 1 as follows:
\begin{equation}\label{scaled_fitness}
  \bar{f}_{\mathrm{fit},j}=\frac{f_{j}^{\mathrm{list}}}{\sum_{s=1}^{N_{\mathrm{tot}}}{f_{s}^{\mathrm{list}}}},
  \;\;\;  j=1,2,...,N_{\mathrm{tot}},
\end{equation}
where $f_{i}^{\mathrm{list}}$ is the fitness value of the $i$th individual in LIST 1.
Based on \eqref{scaled_fitness}, the selection algorithm is given in \emph{Algorithm} \ref{selection_ALG}, which is used to select one parent once at a time.

\begin{algorithm}   
\caption{Algorithm for crossover} 
\label{crossover_ALG} 
\begin{algorithmic}[1] 
\State Generate $c_1$ and $c_2$ from a uniform distribution in $\left( 0,1 \right)$;

    \If{$c_1>p_{\mathrm{c}}$}
       \State $\mathrm{child}1=\mathrm{parent}1$;
       \State $\mathrm{child}2=\mathrm{parent}2$;
    \Else
       \State Set $c_{\mathrm{p}}=\lceil c_2N \rceil$;
       \State $\mathrm{child}1=\left[ \mathrm{parent}1\left( 1:c_{\mathrm{p}} \right) ,\mathrm{parent}2\left( c_{\mathrm{p}}+1:N \right) \right]$;
       \State $\mathrm{child}2=\left[ \mathrm{parent}2\left( 1:c_{\mathrm{p}} \right) ,\mathrm{parent}1\left( c_{\mathrm{p}}+1:N \right) \right]$;
    \EndIf

\end{algorithmic}
\end{algorithm}

\emph{4) Crossover:}
The crossover operation needs two parents. Simply, we execute the selection operation two times to generate two parents for one crossover operation. The crossover algorithm is given in \emph{Algorithm} \ref{crossover_ALG}, where $p_{\mathrm{c}}$ stands for the crossover probability, and $\mathrm{parent}1$ and $\mathrm{parent}2$ are generated from the selection operation.

\begin{algorithm}   
\caption{Algorithm for mutation} 
\label{mutation_ALG} 
\begin{algorithmic}[1] 
\For{$i=1:N$}
    \State Generate $c$ from a uniform distribution in $\left( 0,1 \right)$;
    \If{$c<p_{\mathrm{m}}$}
       \State Reset $\mathrm{child}\left( i \right)$ from a uniform distribution in $\left[ 0,2\pi \right)$;
    \EndIf
\EndFor
\end{algorithmic}
\end{algorithm}

\emph{5) Mutation:}
The mutation operation is after the crossover operation. Each gene of the child generated from the crossover operation can mutate under the mutation probability $p_{\mathrm{m}}$. The mutation operation is shown in \emph{Algorithm} \ref{mutation_ALG}. Finally, the child after the mutation operation is added to the next generation.

Based on \emph{Algorithms} \ref{selection_ALG}-\ref{mutation_ALG}, we propose a GA for the phase shift optimization problem \eqref{phase_opt_continuous} to obtain the optimal phase shifts $\left\{ \theta _{n}^{\mathrm{opt}}\left| n=1,2,...N \right. \right\}$ along with the maximum sum rate $\tilde{R}_{\mathrm{sum}}^{\max}$. The GA is given in \emph{Algorithm} \ref{GA_ALG}.
Additionally, the complexity of the proposed GA-based algorithm is $N_{\mathrm{tot}} \ast n$, where $N_{\mathrm{tot}}$ is the population size, and $n$ is the number of generations evaluated which is determined by the convergence behavior of the GA \cite{8269405}.

\begin{algorithm}   
\caption{GA for phase shift optimization} 
\label{GA_ALG} 
\begin{algorithmic}[1] 
\State \textbf{initialization:} Generate the initial population $\mathrm{P}_1$ with $N_{\mathrm{tot}}$ individuals. The $N$ genes of each individual are initially generated in $\left[ 0,2\pi \right)$. Set the iteration number $t=1$. The number of the termination iteration times is $t_T$, and the termination value is $f_T$.

\Repeat
    \State Calculate the fitness values for $\mathrm{P}_t$ as $\left\{ f_{\mathrm{fit},i}\left| f_{\mathrm{fit},i}=\tilde{R}_{\mathrm{sum},i},i=1,2,...,N_{\mathrm{tot}} \right. \right\}$;
    \State Sort the individuals in $\mathrm{P}_t$, select the top $N_{\mathrm{e}}$ individuals as elites, and add them to the population $\mathrm{P}_{t+1}$;
    \State Scale the fitness values of the individuals in LIST 1;

    \For{$j=1: \left( N_{\mathrm{tot}}-N_{\mathrm{e}}\right)/2 $}
       \State Execute Algorithm 1 two times for selection, and generate two parents as $\mathrm{parent}1$ and $\mathrm{parent}2$;
       \State Execute Algorithm 2 for crossover, and generate two children as $\mathrm{child}1$ and $\mathrm{child}2$;
       \State For each child, execute Algorithm 3 for mutation, and then add $\mathrm{child}1$ and $\mathrm{child}2$ to the population $\mathrm{P}_{t+1}$;
    \EndFor

    \State Set $t=t+1$;
\Until{$t> t_T$ or $\max \left( \left\{ f_{\mathrm{fit},i} \right\} \right) >f_T$}

\State Output $\max \left( \left\{ f_{\mathrm{fit},i} \right\} \right)$ as the maximum sum rate $\tilde{R}_{\mathrm{sum}}^{\max}$ and genes of the corresponding individual as the optimal phase shifts $\left\{ \theta _{n}^{\mathrm{opt}}\left| n=1,2,...N \right. \right\}$.

\end{algorithmic}
\end{algorithm}

It is noted that the phase shifts vary in a continuous range of $\left[ 0,2\pi \right)$ in Problem \eqref{phase_opt_continuous}. However, in practical scenarios, the phase shifts are usually discrete values in the range of $\left[ 0,2\pi \right)$. In that case, the phase shift optimization problem can be formulated as
\begin{align}\label{phase_opt_discrete}
  \max_{\mathbf{\Phi }} \;\;\;  & \tilde{R}_{\mathrm{sum}}   \nonumber \\
  \mathrm{s}.\mathrm{t}.\;\;\;  & \theta _n\in \left\{ 0,\frac{2\pi}{2^B},2\times \frac{2\pi}{2^B},...,\left( 2^B-1 \right) \times \frac{2\pi}{2^B} \right\} ,\; \forall n ,
\end{align}
where the range of the phase shifts is divided into $B$ bits.
It is readily to be seen that with a larger $B$, the phase shifts can be adjusted more accurately.
When the values of the genes are generated from the discrete range, \emph{Algorithm} \ref{GA_ALG} can be applied to solve Problem \eqref{phase_opt_discrete} with discrete phase shifts as to solve Problem \eqref{phase_opt_continuous}.

\section{Numerical Results}
In this section, numerical results are provided.
Similar to the settings in \cite{9090356} and \cite{9743440}, we consider a scenario placed in an XYZ Cartesian coordinate system, where the BS is at coordinate $\left( 0,0,25 \right)$, and the RIS is at $\left( 5,100,30 \right)$.
Users are assumed to be randomly distributed within the circle in Plane $z=1.6$ with the center at $\left( 0,0,1.6 \right)$.
The values for the AoA and AoD of the BS and the RIS are generated from a uniform distribution in $\left( 0,2\pi \right)$.
Furthermore, we set the spacing distance at the BS and the RIS as $d=\frac{\lambda}{2}$,
the number of users as $K=4$,
the Rician factors as $\delta =1$ and $\mu _k=10$, $\forall k$,
the transmit power of users as $p_k=30$ dBm, $\forall k$,
the parameter of the phase noise as $\upsilon = 20$,
the amplitude and phase parameters of RF impairments as $\kappa=0.9$ and $\eta=\pi / 6$,
the noise power as $\sigma ^2=\sigma _{\mathrm{RF}}^{2}=-104$ dBm,
and the ADC quantization bits as $b=2$.
The large-scale fading coefficient is modeled as \vspace{-0.1cm}
\begin{equation}\label{path_loss} \vspace{-0.1cm}
  \mathrm{pathloss}=\frac{l^{-\alpha _{\mathrm{p}}}}{1000},
\end{equation}
where $l$ is the distance between the source and the destination. $\alpha _{\mathrm{p}}$ is the path-loss exponent, which is assumed to be $\alpha _{\mathrm{p}}=2.8$ in this section.
Additionally, the simulation results in this section are obtained by averaging over 2000 Monte Carlo realizations.

\begin{figure}
\vspace{-0.5cm}
  \centering
  \includegraphics[scale=0.55]{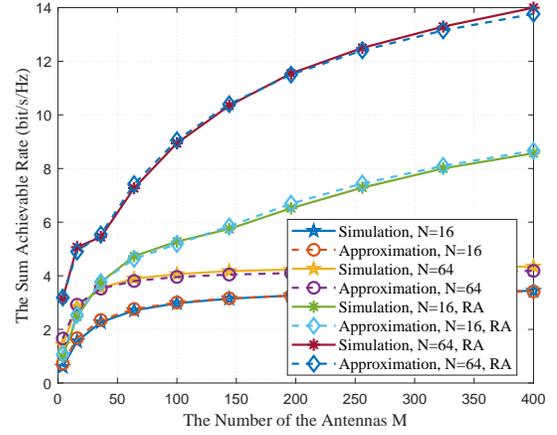}\\  
  \caption{Sum AR versus the number of antennas $M$}\label{p1}  
\end{figure}

In Fig. \ref{p1}, we study how the sum rates vary with the number of the antennas $M$ at the BS, setting the number of the reflecting elements $N$ at the RIS as $N=16$ and $N=64$.
The curves marked with ``Simulation" are obtained based on \eqref{EAR_user_k}, while the curves marked with ``Approximation" are obtained according to \eqref{EAR_user_k_approx} in \emph{Theorem} \ref{theorem_Rk_approx}.
Two cases are considered:
1) \emph{Case} 1: the phase shifts at the RIS are fixed, each of which is randomly generated from a uniform distribution in $\left( 0,2\pi \right)$;
2) \emph{Case} 2: the phase shifts at the RIS are obtained by applying the GA in \emph{Algorithm} \ref{GA_ALG} to Problem \eqref{phase_opt_continuous}.
The parameters of the GA are set as:
the population is $N_{\mathrm{tot}}=200$,
the number of elites is $N_{\mathrm{e}}=10$,
the crossover probability is $p_{\mathrm{c}}=0.4$,
the mutation probability is $p_{\mathrm{m}}=0.1$,
and the number of the termination iteration times is $t_T=2000$.
It can be seen that the approximation curves match well with the simulation curves, which supports the results in \emph{Theorem} \ref{theorem_Rk_approx}.
Moreover, it is noted that the sum rates with the optimized phase shifts obtained by \emph{Algorithm} \ref{GA_ALG} are better than that with the fixed phase shifts.
Furthermore, the sum rates increase with $M$ in Case 2 with the optimized phase shifts, while for Case 1 with the fixed phase shifts, the sum rates first increase and then tend to be saturated with $M$.
Additionally, the sum rates with $N=64$ are higher than that with $N=16$ in both Case 1 and 2.
Therefore, we can improve the performance of the RIS-assisted multi-user massive MIMO system by increasing $M$ and $N$, while using the optimized phase shifts obtained by the proposed GA in \emph{Algorithm} \ref{GA_ALG}.

\begin{figure}
\vspace{-0.5cm}
  \centering
  \includegraphics[scale=0.55]{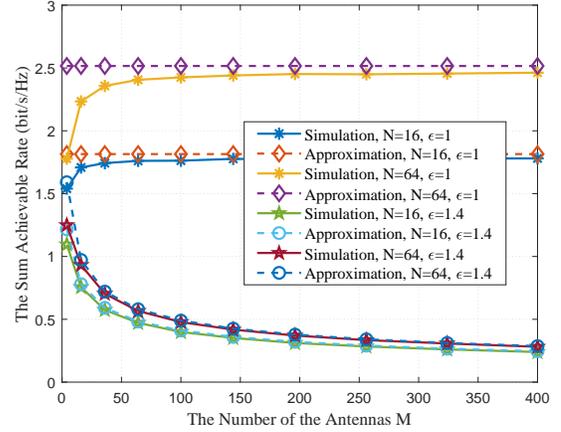}\\  
  \caption{Power scaling laws of the users}\label{p2}  
\end{figure}

\begin{figure}
  \centering
  \includegraphics[scale=0.55]{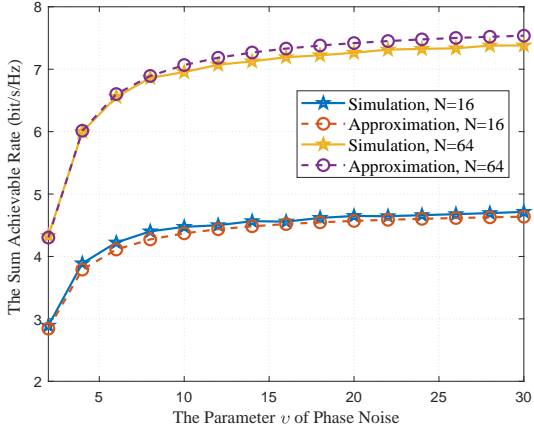}\\  
  \caption{Sum AR versus the parameter $\upsilon$ of RIS phase noise}\label{p3} 
\end{figure}

In Fig. \ref{p2}, we investigate the power scaling laws of the users.
Two groups of the fixed phase shifts are considered, which are obtained by using the GA in \emph{Algorithm} \ref{GA_ALG} with $N=16$ and $N=64$ respectively, when $M=64$.
We set the power as $p_k=\frac{E_u}{M^{\epsilon}}$, $\forall k$, with fixed $E_u = 10$ dB.
The curves marked with ``Approximation" in Fig. \ref{p2} are obtained according to \eqref{Rk_scaling_M}.
It can be seen that when the power scaling factor $\epsilon = 1.4$, as $M\rightarrow \infty$, the sum rates converge to zero in both cases with $N=16$ and $N=64$.
This is because the transmit power is scaled down too fast with $M$.
Furthermore, when $\epsilon = 1$, as $M\rightarrow \infty$, the sum rates converge to non-zero limits in both cases with $N=16$ and $N=64$.
It means the system maintains a required performance when the transmit power is scaled down by the factor $\frac{1}{M}$, which validates our analysis in Section III.

\begin{figure}
\vspace{-0.5cm}
  \centering
  \includegraphics[scale=0.55]{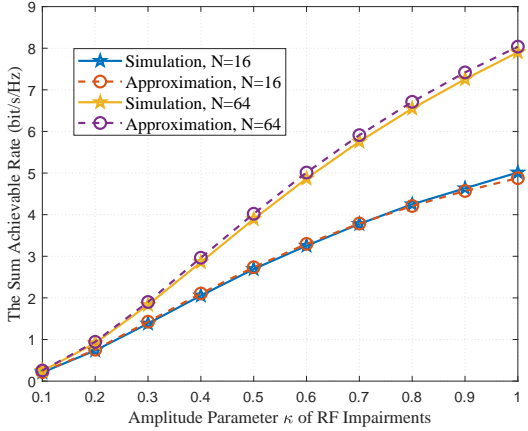}\\  
  \caption{Sum AR versus the amplitude parameter $\kappa$ of RF impairments}\label{p4}  
\end{figure}

\begin{figure}
  \centering
  \includegraphics[scale=0.55]{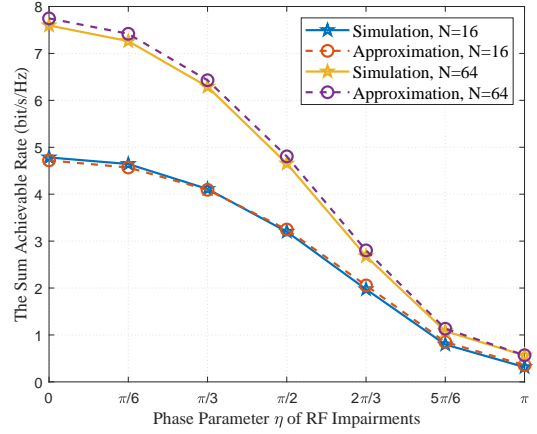}\\  
  \caption{Sum AR versus the phase parameter $\eta$ of RF impairments}\label{p5}  
\end{figure}

Fig. \ref{p3} shows the impacts of the phase noise at the RIS on the sum rates.
It can be seen that the sum rates increase as the concentration parameter $\upsilon$ increases. For both $N=16$ and $N=64$ cases, the increase of the sum rates is rapid when $\upsilon$ is small, and then slows down when $\upsilon$ is large.
Since the phase noise ${\varepsilon _n}$ at the unit $n$ of the RIS follows a Von Mises distribution, a higher concentration parameter $\upsilon$ means the fluctuation for ${\varepsilon _n}$ lies in a smaller range.
Specifically, when $\upsilon \rightarrow \infty$, we have ${\varepsilon _n} \rightarrow 0$, which means there is no phase noise at the RIS.
Therefore, the sum rates converge to the case without phase noise as the concentration parameter $\upsilon$ grows to infinity, which is consistent with the results in Fig. \ref{p3}, where the sum rates first increase and then become saturated as $\upsilon$ increases.

Fig. \ref{p4} and \ref{p5}  focus on the impacts of RF impairments on the sum rates while setting $M=64$.
In Fig. \ref{p4}, we can find that the sun rates increase with the amplitude parameter $\kappa$ in both $N=16$ and $N=64$ cases.
It should be mentioned that $\kappa = 1$ means RF impairments have no impacts on the amplitude of the receiving signals.
Thus, a smaller $\kappa$ stands for severer RF impairments, causing larger reduction of the system performance, as is shown in Fig. \ref{p4}.
In Fig. \ref{p5}, it is observed that the sun rates decrease with the phase parameter $\eta$ in both $N=16$ and $N=64$ cases.
This is because when $\eta$ is large, the phase shifts brought by RF impairments would lie in a large range, which leads to poor system performance.

\begin{figure}
\vspace{-0.5cm}
  \centering
  \includegraphics[scale=0.55]{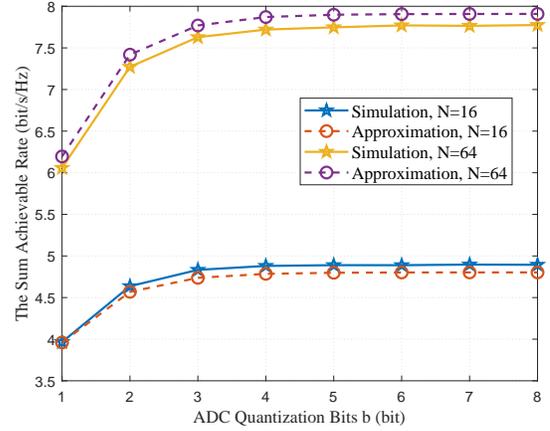}\\  
  \caption{Sum AR versus ADC quantization bits $b$}\label{p6}  
\end{figure}

\begin{figure}
  \centering
  \includegraphics[scale=0.55]{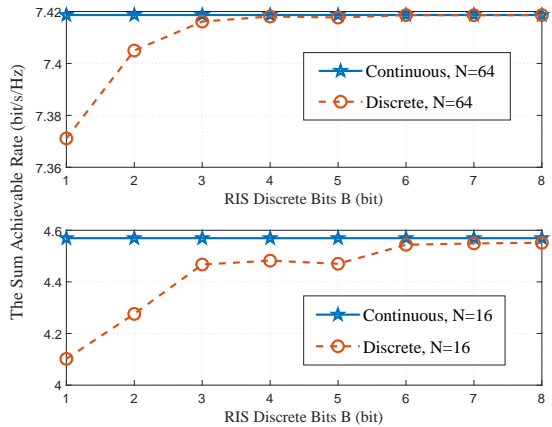}\\  
  \caption{The comparison of the continuous and the discrete phase shifts}\label{p7}  
\end{figure}

Fig. \ref{p6} depicts how the sum rates vary with the ADC quantization bits $b$.
It is readily seen that in both $N=16$ and $N=64$ cases, the sum rates first increase fast with $b$, and then become saturated.
It is well known that the power consumption and hardware cost of wireless systems increase rapidly as ADC quantization bits $b$ increases..
Thus, Fig. \ref{p6} indicates that we can choose $b=3$ for the considered system to obtain a balance between system performance, power consumption and hardware cost.

Finally, we make a comparison between the cases with continuous phase shifts and with discrete phase shifts in Fig. \ref{p7}.
The phase shifts for the curves marked with "Discrete" is obtained by applying \emph{Algorithm} \ref{GA_ALG} to Problem \eqref{phase_opt_discrete} with discrete phase shifts.
The RIS discrete bits $B$ is defined in \eqref{phase_opt_discrete}.
In practical scenarios, discrete phase shifts are used, while the continuous phase shifts are considered for ideal scenarios.
Therefore, it is essential to study the system performance with discrete phase shifts.
From Fig. \ref{p7}, the sum rates with the optimized discrete phase shifts increase as $B$ increases, and converge to that with the optimized continuous phase shifts.
It is noted that $B=6$ and $B=3$ are reasonable choices for the discrete case with $N=16$ and $N=64$, respectively, since they achieve the performance for the corresponding continuous cases, while a higher $B$ leads to more cost.

\section{Conclusion}
The paper studied an RIS-assisted multi-user uplink massive MIMO system under Rician fading channels and with imperfect hardware at both the RIS and BS.
At the RIS, the paper used Von Mises distribution to model the phase noise, while at the BS, the paper adopted EEVM model for RF impairments and AQNM for low-resolution ADCs.
Based on that, asymptotic data rates were obtained with infinite $M$ and $N$, when the RIS was aligned to a specific user and when the RIS was aligned to none of the users.
Besides, the paper investigated the power scaling laws of the users, and showed that while guaranteeing a required system performance, the transmit power of the users can be scaled down at most by the factor $\frac{1}{M}$ when $M$ goes infinite, or by the factor $\frac{1}{MN}$ when $M$ and $N$ go infinite.
Furthermore, an optimization algorithm was proposed based on GA, which can be applied to solve the continuous and discrete phase shift optimization problems for maximizing the sum rates of the system.
Numerical results were provided to support the main results.
The numerical results revealed that the sum rate of the considered system can be improved by increasing $M$ and $N$.
Moreover, the impacts of the parameters of the imperfect hardware on the system performance were revealed.
The numerical results also revealed that when $M=64$, $B=6$ and $B=3$ are reasonable choices for the discrete cases with $N=16$ and $N=64$, respectively.

\begin{figure*}[t]
\setcounter{equation}{44}
\begin{equation}\label{EAR_user_k_approx_afterlemma1}
  R_k\approx \tilde{R}_k=\log _2\left( 1+\frac{p_k\tau ^2 \mathrm{E}\left\{\left| \mathbf{h}_{k}^{H}\mathbf{\Phi }^H\mathbf{G}^H\bm{\chi} \mathbf{G\Phi \Theta h}_k \right|^2 \right\}}{ \sum_{i\ne k}^K{p_i\tau ^2\mathrm{E}\left\{\left| \mathbf{h}_{k}^{H}\mathbf{\Phi }^H\mathbf{G}^H\bm{\chi} \mathbf{G\Phi \Theta h}_i \right|^2\right\}} +\mathrm{E}\left\{ \mathrm{DN}_k \right\} +\mathrm{E}\left\{ \mathrm{AN}_k \right\} +\mathrm{E}\left\{ \mathrm{QN}_k \right\}} \right)
\end{equation}
\setcounter{equation}{37}
\hrulefill 

\setcounter{equation}{45}  \vspace{-0.2cm}
\begin{align}\label{hk_hk}
  &\mathbf{h}_{k}^{H}\mathbf{\Phi }^H\mathbf{G}^H\bm{\chi} \mathbf{G\Phi \Theta h}_{k}^{} = \mathbf{h}_{k}^{H}\mathbf{\Phi }^H\frac{\beta}{\delta +1}\left( \delta \mathbf{\bar{G}}^H\bm{\chi}\mathbf{\bar{G}} + \sqrt{\delta}\mathbf{\bar{G}}^H\bm{\chi}\mathbf{\tilde{G}} + \sqrt{\delta}\mathbf{\tilde{G}}^H\bm{\chi}\mathbf{\bar{G}} + \mathbf{\tilde{G}}^H\bm{\chi}\mathbf{\tilde{G}} \right) \mathbf{\Phi \Theta h}_k  \nonumber \\
  &=\frac{\beta \alpha _k}{\left( \delta +1 \right) \left( \mu _k+1 \right)}\left( \sqrt{\mu _k}\mathbf{\bar{h}}_{k}^{H}+\mathbf{\tilde{h}}_{k}^{H} \right) \mathbf{A\Theta }\left( \sqrt{\mu _k}\mathbf{\bar{h}}_{k}^{}+\mathbf{\tilde{h}}_{k}^{} \right)  \nonumber \\
  &=\frac{\beta \alpha _k}{\left( \delta +1 \right) \left( \mu _k+1 \right)}( {\underbrace{\mu _k\mathbf{\bar{h}}_{k}^{H}\mathbf{A\Theta \bar{h}}_{k}^{}}_{\omega _{kk}^{1}}} +  {\underbrace{\sqrt{\mu _k}\mathbf{\bar{h}}_{k}^{H}\mathbf{A\Theta \tilde{h}}_{k}^{}}_{\omega _{kk}^{2}}} +  {\underbrace{\sqrt{\mu _k}\mathbf{\tilde{h}}_{k}^{H}\mathbf{A\Theta \bar{h}}_{k}^{}}_{\omega _{kk}^{3}}} +  {\underbrace{\mathbf{\tilde{h}}_{k}^{H}\mathbf{A\Theta \tilde{h}}_{k}^{}}_{\omega _{kk}^{4}}})
\end{align}
\setcounter{equation}{37}
\hrulefill 
\end{figure*}


\begin{figure*}[b]
\vspace{-0.01cm}
\hrulefill 
\setcounter{equation}{50}  \vspace{-0.1cm}
\begin{equation}\label{wkk_12} \vspace{-0.2cm}
  \hspace{-3cm} \omega _{kk}^{1,2}=\kappa f_{k}^{\ast}\sum_{n=1}^N{\sum_{m=1}^M{a_{Mm}^{*}\left( \phi _{r}^{a},\phi _{r}^{e} \right) e^{j\varphi _m}\tilde{g}_{mn}e^{j\theta _n}e^{j\varepsilon _n}a_{Nn}^{}\left( \phi _{kr}^{a},\phi _{kr}^{e} \right)}}
\end{equation}
\begin{equation}\label{wkk_13} \vspace{-0.2cm}
  \hspace{-2cm} \omega _{kk}^{1,3}=\kappa \sum_{n=1}^N{\sum_{m=1}^M{a_{Mm}^{}\left( \phi _{r}^{a},\phi _{r}^{e} \right) e^{j\varphi _m}\tilde{g}_{mn}^{*}e^{-j\theta _n}a_{Nn}^{*}\left( \phi _{kr}^{a},\phi _{kr}^{e} \right)}}\sum_{s=1}^N{e^{j\varepsilon _s}f_{k,s}}
\end{equation}
\begin{equation}\label{wkk_14}
  \omega _{kk}^{1,4}=\kappa \sum_{m=1}^M{e^{j\varphi _m}\sum_{n_1=1}^N{a_{Nn_1}^{*}\left( \phi _{kr}^{a},\phi _{kr}^{e} \right) e^{-j\theta _{n_1}}\tilde{g}_{mn_1}^{*}}\sum_{n_2=1}^N{\tilde{g}_{mn_2}^{}e^{j\theta _{n_2}}e^{j\varepsilon _{n_2}}a_{Nn_2}^{}\left( \phi _{kr}^{a},\phi _{kr}^{e} \right)}}
\end{equation}
\setcounter{equation}{37}
\end{figure*}

It should be aware that the considered system model can be further extended to investigate a more practical scenario. For example, this work considered perfect CSI, while perfect CSI is usually hard to obtain in practical.
Furthermore, due to the large number of antennas at BS and reflecting units at RIS, channels are usually correlated with each other in a practical RIS-assisted massive MIMO system. Therefore, the future work would take channel estimation and channel correlation into consideration to provide more insights for RIS-assisted massive MIMO systems.

\begin{appendices}

\section{Proof of \emph{Theorem}~\ref{theorem_Rk_approx}}\label{derivation_Rk_approx}
First, we review some key preliminary results given in the following lemmas.

\emph{Lemma 1:} \cite[\emph{Lemma 1}]{6816003} If  $X = \sum\nolimits_{i = 1}^{{t_1}} {{X_i}} $ and $Y = \sum\nolimits_{j = 1}^{{t_2}} {{Y_j}} $  are both sums of nonnegative random variables ${X_i}$  and  ${Y_j}$, then we get the following approximation  \vspace{-0.05cm}
\begin{equation}\label{9-1}  \vspace{-0.05cm}
{\rm E}\left\{ {{{\log }_2}\left( {1 + \frac{X}{Y}} \right)} \right\} \approx {\log _2}\left( {1 + \frac{{{\rm E}\left\{ X \right\}}}{{{\rm E}\left\{ Y \right\}}}} \right).
\end{equation}
Note that it is not necessary for the random variables  $X$ and $Y$  to be independent.
In addition, the approximation becomes more accurate as ${t_1}$  and  ${t_2}$ increase.

\emph{Lemma 2:} If $\varepsilon$ is a random variable following zero-mean Von Mises distribution with a concentration parameter $\upsilon$, the characteristic function of $\varepsilon$, $\mathrm{E}\left\{ e^{jt\varepsilon} \right\}$, equals to $\frac{I_1\left( \nu \right)}{I_0\left( \nu \right)}$ when $t = 1$.
In other words, we have  \vspace{-0.1cm}
\begin{equation}\label{rho_value} \vspace{-0.1cm}
  \mathrm{E}\left\{ e^{j\varepsilon} \right\} =\frac{I_1\left( \upsilon \right)}{I_0\left( \upsilon \right)} \triangleq \rho ,
\end{equation}
where $I_n\left( \upsilon \right)$ represents the modified Bessel function of the first kind and order $n$.

\begin{IEEEproof}
The Bessel Function of the first kind and order $n$ is given by \vspace{-0.1cm}
\begin{equation}\label{B1_n} \vspace{-0.05cm}
  J_n\left( \upsilon \right) =\frac{1}{2\pi}\int_0^{2\pi}{e^{j\left( n\varepsilon -\upsilon \sin \varepsilon \right)}d\varepsilon},
\end{equation}
and the corresponding Modified Bessel Function is given by  \vspace{-0.1cm}
\begin{equation}\label{mB1_n}  \vspace{-0.1cm}
  I_n\left( \upsilon \right) =j^{-n}J_n\left( j\upsilon \right) =j^{-n}\frac{1}{2\pi}\int_0^{2\pi}{e^{jn\varepsilon +\upsilon \sin \varepsilon}d\varepsilon}.
\end{equation}
When $n=1$, we can obtain 
\begin{align}\label{mB1_1}
  & I_1\left( \upsilon \right) = j^{-1}J_1\left( j\upsilon \right) = j^{-1}\frac{1}{2\pi} \! \int_0^{2\pi}{e^{j\varepsilon +\upsilon \sin \varepsilon}d\varepsilon} \nonumber \\
  & = e^{-j\frac{\pi}{2}}\frac{1}{2\pi} \! \int_0^{2\pi}{e^{j\varepsilon +\upsilon \sin \varepsilon}d\varepsilon}  \nonumber \\
  & = \frac{1}{2\pi} \! \int_0^{2\pi}{e^{j\left( \varepsilon -\frac{\pi}{2} \right) +\upsilon \sin \left( \varepsilon -\frac{\pi}{2}+\frac{\pi}{2} \right)}d\varepsilon}  \nonumber \\
  & = \frac{1}{2\pi} \! \int_0^{2\pi}{e^{j\left( \varepsilon -\frac{\pi}{2} \right) +\upsilon \cos \left( \varepsilon -\frac{\pi}{2} \right)}d\varepsilon} = \frac{1}{2\pi} \! \int_0^{2\pi}{e^{j\varepsilon +\upsilon \cos \varepsilon}d\varepsilon}.
\end{align}

On the other hand, the PDF of $\varepsilon$ is given by \vspace{-0.1cm}
\begin{equation}\label{pdf_VMs} \vspace{-0.1cm}
  M_{\upsilon}\left( \varepsilon \right) =\frac{1}{2\pi I_0\left( \upsilon \right)}e^{\upsilon \cos \varepsilon},\;\;\; \varepsilon \in \left[ 0,2\pi \right).
\end{equation}
Then the characteristic function with $t=1$ can be calculated as \vspace{-0.1cm}
\begin{equation*} \vspace{-0.2cm}
  \mathrm{E}\left\{ e^{j\varepsilon} \right\} =\int_0^{2\pi}{e^{j\varepsilon}M_{\upsilon}\left( \varepsilon \right) d\varepsilon}=\int_0^{2\pi}{e^{j\varepsilon}\frac{1}{2\pi I_0\left( \upsilon \right)}e^{\upsilon \cos \varepsilon}d\varepsilon}
\end{equation*}
\begin{equation}\label{cf1_VMs} \vspace{-0.1cm}
  \hspace{-2cm} =\frac{1}{2\pi I_0\left( \upsilon \right)}\int_0^{2\pi}{e^{j\varepsilon +\upsilon \cos \varepsilon}d\varepsilon}=\frac{I_1\left( \upsilon \right)}{I_0\left( \upsilon \right)}=\rho.
\end{equation}
\end{IEEEproof}

From \eqref{EAR_user_k}, by using \emph{Lemma 1}, we can obtain \eqref{EAR_user_k_approx_afterlemma1} at the top of this page. \addtocounter{equation}{1}
Then, we start to derive the expectations in \eqref{EAR_user_k_approx_afterlemma1} one by one.

Based on \eqref{channel_H} and \eqref{channel_G}, the term $\mathbf{h}_{k}^{H}\mathbf{\Phi }^H\mathbf{G}^H\bm{\chi} \mathbf{G\Phi \Theta h}_k$ can be written as \eqref{hk_hk} at the top of this page, \addtocounter{equation}{1}
where the matrix $\mathbf{A}$ is defined as \vspace{-0.1cm}
\begin{equation}\label{A_defination} \vspace{-0.1cm}
  \mathbf{A} = \mathbf{\Phi }^H \! \left( \delta \mathbf{\bar{G}}^H\bm{\chi} \mathbf{\bar{G}} \!+\! \sqrt{\delta}\mathbf{\bar{G}}^H\bm{\chi} \mathbf{\tilde{G}} \!+\! \sqrt{\delta}\mathbf{\tilde{G}}^H\bm{\chi} \mathbf{\bar{G}} \!+\! \mathbf{\tilde{G}}^H\bm{\chi} \mathbf{\tilde{G}} \right) \! \mathbf{\Phi }.
\end{equation}
Therefore, $\mathrm{E}\left\{\left| \mathbf{h}_{k}^{H}\mathbf{\Phi }^H\mathbf{G}^H\bm{\chi} \mathbf{G\Phi \Theta h}_k \right|^2 \right\}$ can be expressed as \vspace{-0.1cm}
\begin{equation*} \vspace{-0.2cm}
  \hspace{-4.3cm} \mathrm{E}\left\{\left| \mathbf{h}_{k}^{H}\mathbf{\Phi }^H\mathbf{G}^H\bm{\chi} \mathbf{G\Phi \Theta h}_k \right|^2 \right\}
\end{equation*}
\begin{equation*} \vspace{-0.2cm}
  \hspace{-3cm} =\frac{\beta ^2\alpha _{k}^{2}}{\left( \delta +1 \right) ^2\left( \mu _k+1 \right) ^2}\mathrm{E}\left\{ \left| \sum_{i=1}^4{\omega _{kk}^{i}} \right|^2 \right\}
\end{equation*}
\begin{equation*} \vspace{-0.2cm}
  \hspace{-5.4cm} =\frac{\beta ^2\alpha _{k}^{2}}{\left( \delta +1 \right) ^2\left( \mu _k+1 \right) ^2}
\end{equation*}
\begin{equation*} \vspace{-0.2cm}
  \times \left( \sum_{i=1}^4{\mathrm{E}\left\{ \left| \omega _{kk}^{i} \right|^2 \right\}}+2\sum_{i=1}^4{\sum_{j=i+1}^4{\mathrm{E}\left\{ \mathrm{Re}\left( \omega _{kk}^{i}\left( \omega _{kk}^{j} \right) ^* \right) \right\}}} \right)
\end{equation*}
\begin{equation*} \vspace{-0.2cm}
  \hspace{-5.5cm} \overset{\left( a \right)}{=}\frac{\beta ^2\alpha _{k}^{2}}{\left( \delta +1 \right) ^2\left( \mu _k+1 \right) ^2}
\end{equation*}
\begin{equation}\label{hk_hk_norm2_expand} \vspace{-0.1cm}
  \hspace{-1.4cm} \times \left( \sum_{i=1}^4{\mathrm{E}\left\{ \left| \omega _{kk}^{i} \right|^2 \right\}}+2\mathrm{E}\left\{ \mathrm{Re}\left( \omega _{kk}^{1}\left( \omega _{kk}^{4} \right) ^* \right) \right\} \right),
\end{equation}
where step $\left( a \right)$ is obtained by removing the zero-expectation terms.
Then, we focus on deriving the expectation $\mathrm{E}\left\{ \left| \omega _{kk}^{1} \right|^2 \right\}$. Note that \vspace{-0.1cm}
\begin{align}\label{w1_kk_expand}
  & \omega _{kk}^{1}  = \mu _k\mathbf{\bar{h}}_{k}^{H}\mathbf{A\Theta \bar{h}}_{k}^{}   \nonumber \\
  & = \delta \mu _k {\underbrace{\mathbf{\bar{h}}_{k}^{H}\mathbf{\Phi }^H\mathbf{\bar{G}}^H \bm{\chi} \mathbf{\bar{G}\Phi \Theta \bar{h}}_{k}^{}}_{\omega _{kk}^{1,1}}} + \sqrt{\delta}\mu _k {\underbrace{\mathbf{\bar{h}}_{k}^{H}\mathbf{\Phi }^H\mathbf{\bar{G}}^H \bm{\chi} \mathbf{\tilde{G}\Phi \Theta \bar{h}}_{k}^{}}_{\omega _{kk}^{1,2}}}   \nonumber \\
  & + \sqrt{\delta}\mu _k {\underbrace{\mathbf{\bar{h}}_{k}^{H}\mathbf{\Phi }^H\mathbf{\tilde{G}}^H \bm{\chi} \mathbf{\bar{G}\Phi \Theta \bar{h}}_{k}^{}}_{\omega _{kk}^{1,3}}} + \mu _k {\underbrace{\mathbf{\bar{h}}_{k}^{H}\mathbf{\Phi }^H\mathbf{\tilde{G}}^H \bm{\chi} \mathbf{\tilde{G}\Phi \Theta \bar{h}}_{k}^{}}_{\omega _{kk}^{1,4}}}   \nonumber \\
  & =\delta \mu _k\omega _{kk}^{1,1}+\sqrt{\delta}\mu _k\omega _{kk}^{1,2}+\sqrt{\delta}\mu _k\omega _{kk}^{1,3}+\mu _k\omega _{kk}^{1,4},
\end{align}
where the term $\omega _{kk}^{1,1}$ in \eqref{w1_kk_expand} is expressed as \vspace{-0.1cm}
\begin{equation}\label{wkk_11} \vspace{-0.1cm}
  \omega _{kk}^{1,1}=\kappa f_{k}^{\ast}\sum_{m=1}^M{e^{j\varphi _m}}\sum_{n=1}^N{e^{j\varepsilon _n}f_{k,n}},
\end{equation}
and $\omega _{kk}^{1,2}$, $\omega _{kk}^{1,3}$ and $\omega _{kk}^{1,4}$ in \eqref{w1_kk_expand} are respectively expressed as \eqref{wkk_12}-\eqref{wkk_14} at the bottom of the page, \addtocounter{equation}{3}
where the terms $f_k$ and $f_{k,n}$ are defined as \vspace{-0.1cm}
\begin{equation}\label{fk} \vspace{-0.1cm}
   f_k \! \triangleq \! \sum_{n=1}^N{f_{k,n}}, \;
   f_{k,n} \! \triangleq \! a_{Nn}^{*} \! \left( \phi _{t}^{a},\phi _{t}^{e} \right) e^{j\theta _n}a_{Nn} \! \left( \phi _{kr}^{a},\phi _{kr}^{e} \right).
\end{equation}
\begin{figure*}[b]
\hrulefill 
\setcounter{equation}{61}
\begin{equation*} \vspace{-0.2cm}
  \mathrm{E}\left\{ \left| \omega _{kk}^{1} \right|^2 \right\} = \kappa ^2M \Big( \left( \left( \rho ^2N+\left( 1-\rho ^2 \right) \delta ^2\left| f_k \right|^2+1-\rho ^2+2\rho ^2\delta \left| f_k \right|^2 \right) N+\rho ^2\delta ^2\left| f_k \right|^4+2\left( 1-\rho ^2 \right) \delta \left| f_k \right|^2 \right) \iota ^2\mu _{k}^{2}M
\end{equation*}
\begin{equation*} \vspace{-0.2cm}
  +\left( \left( 1-\rho ^2 \right) \delta +\left( 1+\rho ^2-\iota ^2\rho ^2 \right) \right) \mu _{k}^{2}N^2 +2\left( 1-\iota ^2 \right) \left( 1-\rho ^2 \right) \delta \mu _{k}^{2}\left| f_k \right|^2+\left( 1-\iota ^2 \right) \rho ^2\delta ^2\mu _{k}^{2}\left| f_k \right|^4
\end{equation*}
\begin{equation}\label{w1_kk_norm2_value} \vspace{-0.1cm}
  +\left( \left( 1-\iota ^2 \right) \left( 1-\rho ^2 \right) \delta ^2\left| f_k \right|^2+\delta \left| f_k \right|^2+\delta \rho ^2\left| f_k \right|^2+\left( 1-\rho ^2 \right) \left( 1-\iota ^2 \right) +2\left( 1-\iota ^2 \right) \rho ^2\delta \left| f_k \right|^2 \right) \mu _{k}^{2}N \Big)
\end{equation}
\begin{equation*} \vspace{-0.2cm}
  \mathrm{E}\left\{ \left| \omega _{kk}^{2} \right|^2 \right\} = \kappa ^2M \Big( \left( \left( \delta ^2\left| f_k \right|^2+1 \right) N+2\delta \left| f_k \right|^2 \right) \iota ^2\mu _kM+\left( \delta +1 \right) \mu _kN^2
\end{equation*}
\begin{equation}\label{w2_kk_norm2_value} \vspace{-0.1cm}
  +\left( \left( 1-\iota ^2 \right) \delta ^2\left| f_k \right|^2+\delta \left| f_k \right|^2+1-\iota ^2 \right) \mu _kN+2\left( 1-\iota ^2 \right) \delta \mu _k\left| f_k \right|^2 \Big)
\end{equation}
\begin{equation*} \vspace{-0.2cm}
  \mathrm{E}\left\{ \left| \omega _{kk}^{3} \right|^2 \right\} =\kappa ^2M \Big( \left( \left( 1-\rho ^2 \right) \delta ^2N^2+\left( \delta ^2\rho ^2\left| f_k \right|^2+2\left( 1-\rho ^2 \right) \delta +1 \right) N+2\delta \rho ^2\left| f_k \right|^2 \right) \iota ^2\mu _kM
\end{equation*}
\begin{equation*} \vspace{-0.2cm}
  +\left( \left( 1-\rho ^2 \right) \left( 1-\iota ^2 \right) \delta ^2+\delta +\left( 1-\rho ^2 \right) \delta +1 \right) \mu _kN^2 +2\left( 1-\iota ^2 \right) \delta \rho ^2\mu _k\left| f_k \right|^2
\end{equation*}
\begin{equation}\label{w3_kk_norm2_value} \vspace{-0.1cm}
  +\left( \left( 1-\iota ^2 \right) \delta ^2\rho ^2\left| f_k \right|^2+\delta \rho ^2\left| f_k \right|^2+1-\iota ^2+2\left( 1-\iota ^2 \right) \left( 1-\rho ^2 \right) \delta \right) \mu _kN \Big)
\end{equation}
\begin{equation*} \vspace{-0.2cm}
  \mathrm{E}\left\{ \left| \omega _{kk}^{4} \right|^2 \right\} =\kappa ^2M \Big( \left( \left( \left( \rho ^2+1 \right) \delta ^2+\rho ^2+2\rho ^2\delta \right) N+\left( 1-\rho ^2 \right) \delta ^2+2-\rho ^2+2\left( 2-\rho ^2 \right) \delta \right) \iota ^2MN
\end{equation*}
\begin{equation*} \vspace{-0.2cm}
  +\left( \left( 1-\iota ^2 \right) \left( \rho ^2+1 \right) \delta ^2+2\delta +\rho ^2+1-\iota ^2\rho ^2+2\left( 1-\iota ^2 \right) \rho ^2\delta \right) N^2
\end{equation*}
\begin{equation}\label{w4_kk_norm2_value} \vspace{-0.1cm}
  +\left( \left( 1-\iota ^2 \right) \left( 1-\rho ^2 \right) \delta ^2+2\delta +3-\rho ^2-2\iota ^2+\iota ^2\rho ^2+2\left( 1-\iota ^2 \right) \left( 2-\rho ^2 \right) \delta \right) N \Big)
\end{equation}
\begin{equation*} \vspace{-0.2cm}
  \mathrm{E}\left\{ \mathrm{Re}\left( \omega _{kk}^{1}\left( \omega _{kk}^{4} \right) ^* \right) \right\} =\kappa ^2M \Big( \left( \left( \delta +1 \right) \rho ^2N^2+\left( \rho ^2\delta \left| f_k \right|^2+1-\rho ^2 \right) \left( \delta +1 \right) N+\left( 1-\rho ^2 \right) \delta \left( \delta +1 \right) \left| f_k \right|^2 \right) \iota ^2\mu _kM
\end{equation*}
\begin{equation*} \vspace{-0.2cm}
  +\left( 1-\iota ^2 \right) \left( \delta +1 \right) \rho ^2\mu _kN^2 +\left( \left( 1-\iota ^2 \right) \left( 1-\rho ^2 \right) \left( \delta +1 \right) +1+\rho ^2 \right) \delta \mu _k\left| f_k \right|^2
\end{equation*}
\begin{equation}\label{w1_kk_w4_kk_value}
  +\left( \left( 1-\iota ^2 \right) \rho ^2\delta \left( \delta +1 \right) \left| f_k \right|^2+\left( 2-\iota ^2 \right) \left( 1-\rho ^2 \right) \delta +2-\iota ^2-\rho ^2+\iota ^2\rho ^2 \right) \mu _kN   \Big)
\end{equation}
\setcounter{equation}{54}
\end{figure*}
Thus, the expectation ${\rm E}\left\{ {{{\left| {\omega _{kk}^1} \right|}^2}} \right\}$ can be calculated as \vspace{-0.1cm}
\begin{equation*} \vspace{-0.1cm}
  \hspace{-6.8cm} \mathrm{E}\left\{ \left| \omega _{kk}^{1} \right|^2 \right\}
\end{equation*}
\begin{equation*} \vspace{-0.1cm}
  \hspace{-0.6cm} =\mathrm{E}\left\{ \left| \delta \mu _k\omega _{kk}^{1,1}+\sqrt{\delta}\mu _k\omega _{kk}^{1,2}+\sqrt{\delta}\mu _k\omega _{kk}^{1,3}+\mu _k\omega _{kk}^{1,4} \right|^2 \right\}
\end{equation*}
\begin{equation*} \vspace{-0.1cm}
  \overset{\left( a \right)}{=}\delta ^2\mu _{k}^{2}\mathrm{E}\left\{ \left| \omega _{kk}^{1,1} \right|^2 \right\} +\delta \mu _{k}^{2}\mathrm{E}\left\{ \left| \omega _{kk}^{1,2} \right|^2 \right\} +\delta \mu _{k}^{2}\mathrm{E}\left\{ \left| \omega _{kk}^{1,3} \right|^2 \right\}
\end{equation*}
\begin{equation}\label{w1_kk_norm2} \vspace{-0.1cm}
  +\mu _{k}^{2}\mathrm{E}\left\{ \left| \omega _{kk}^{1,4} \right|^2 \right\} + 2\delta \mu _{k}^{2}\mathrm{E}\left\{ \mathop {\mathrm{Re}}_{}\left( \omega _{kk}^{1,1}\times \left( \omega _{kk}^{1,4} \right) ^* \right) \right\}.
\end{equation}
Step $(a)$ is obtained by removing zero-valued terms. The expectation $\mathrm{E}\left\{ \left| \omega _{kk}^{1,1} \right|^2 \right\}$ in \eqref{w1_kk_norm2} is further calculated as \vspace{-0.1cm}
\begin{equation*} \vspace{-0.1cm}
  \hspace{-1.2cm} \mathrm{E}\left\{ \left| \omega _{kk}^{1,1} \right|^2 \right\} = \mathrm{E}\left\{ \left| \kappa f_{k}^{\ast}\sum_{m=1}^M{e^{j\varphi _m}}\sum_{n=1}^N{e^{j\varepsilon _n}f_{k,n}} \right|^2 \right\}
\end{equation*}
\begin{equation*} \vspace{-0.1cm}
  \hspace{-0.7cm} =\kappa ^2\left| f_k \right|^2\mathrm{E}\left\{ \left| \sum_{m=1}^M{e^{j\varphi _m}} \right|^2 \right\} \mathrm{E}\left\{ \left| \sum_{n=1}^N{e^{j\varepsilon _n}f_{k,n}} \right|^2 \right\}
\end{equation*}
\begin{equation*} \vspace{-0.2cm}
  \hspace{-2.3cm} =\kappa ^2\left| f_k \right|^2\left( M+M\left( M-1 \right) \frac{\sin ^2\left( \eta \right)}{\eta ^2} \right)
\end{equation*}
\begin{equation*} \vspace{-0.1cm}
  \times \left( N+\rho ^2\sum_{n_1=1}^N{\sum_{n_2\ne n_1}^N{f_{k,n_1}\left( f_{k,n_2} \right) ^*}} \right)
\end{equation*}
\begin{equation}\label{wkk_11_norm2} \vspace{-0.1cm}
  =M\kappa ^2\left| f_k \right|^2\left( \iota ^2M+1-\iota ^2 \right) \left( \left( 1-\rho ^2 \right) N+\rho ^2\left| f_k \right|^2 \right),
\end{equation}
where scalar $\iota$ is defined as   \vspace{-0.1cm}
\begin{equation}\label{iota_definition} \vspace{-0.1cm}
  \iota \triangleq \mathrm{E}\left\{ e^{j\varphi _m} \right\} =\frac{\sin \left( \eta \right)}{\eta}=\mathrm{E}\left\{ e^{-j\varphi _m} \right\}.
\end{equation}
The $\rho$ is given by \eqref{rho_value} in \emph{Lemma 2}. Additionally, it is well known that $\mathrm{E}\left\{ e^{-j\varepsilon} \right\} =\mathrm{E}\left\{ e^{j\varepsilon} \right\}$, since $\rho$ is real.
The other expectations in \eqref{w1_kk_norm2} can be obtained in a similar way, which are respectively expressed as follows: \vspace{-0.1cm}
\begin{equation}\label{wkk_12_norm2} \vspace{-0.1cm}
  \hspace{-3.9cm} \mathrm{E}\left\{ \left| \omega _{kk}^{1,2} \right|^2 \right\} = \kappa ^2MN\left| f_k \right|^2,
\end{equation}
\begin{equation}\label{wkk_13_norm2} \vspace{-0.1cm}
  \mathrm{E}\left\{ \left| \omega _{kk}^{1,3} \right|^2 \right\}= \kappa ^2\rho ^2MN\left| f_k \right|^2+\kappa ^2\left( 1-\rho ^2 \right) MN^2,
\end{equation}
\begin{equation*} \vspace{-0.2cm}
  \mathrm{E}\left\{ \left| \omega _{kk}^{1,4} \right|^2 \right\}= \kappa ^2MN\left( \iota ^2M+1-\iota ^2 \right) \left( \rho ^2N+1-\rho ^2 \right)
\end{equation*}
\begin{equation}\label{wkk_14_norm2} \vspace{-0.1cm}
  \hspace{-2.1cm} +\kappa ^2MN^2,
\end{equation}
\begin{equation*} \vspace{-0.1cm}
  \hspace{-0.2cm} \mathrm{E}  \left\{ \mathrm{Re} \! \left( \omega _{kk}^{1,1}  \left( \omega _{kk}^{1,4} \right) ^* \right) \right\} = \kappa ^2\left| f_k \right|^2M\left( \rho ^2N+1-\rho ^2 \right)
\end{equation*}
\begin{equation}\label{wkk_11_wkk_14} \vspace{-0.1cm}
  \hspace{2.3cm} \times \left( \iota ^2M+1-\iota ^2 \right).
\end{equation}
Then, we arrived at \eqref{w1_kk_norm2_value} at the bottom of this page.
Similar to the calculation for ${\rm E}\left\{ {{{\left| {\omega _{kk}^1} \right|}^2}} \right\}$, the rest expectations in \eqref{hk_hk_norm2_expand} can be obtained as \eqref{w2_kk_norm2_value}-\eqref{w1_kk_w4_kk_value} at the bottom of this page.
\addtocounter{equation}{5}
Substituting \eqref{w1_kk_norm2_value}-\eqref{w1_kk_w4_kk_value} into \eqref{hk_hk_norm2_expand}, we can obtain \vspace{-0.1cm}
\begin{equation*} \vspace{-0.1cm}
  \hspace{-1.2cm} \mathrm{E}\left\{ \left| \mathbf{h}_{k}^{H}\mathbf{\Phi }^H\mathbf{G}^H \bm{\chi} \mathbf{G\Phi \Theta h}_{k}^{} \right|^2 \right\} = \frac{\beta ^2\alpha _{k}^{2}\kappa ^2M}{\left( \delta +1 \right) ^2\left( \mu _k+1 \right) ^2}
\end{equation*}
\begin{equation}\label{hk_hk_norm2} \vspace{-0.05cm}
   \times \left( c_{k,1}\iota ^2M+c_{k,2}N^2+c_{k,3}N+c_{k,4} \right) \triangleq \xi _k,
\end{equation}
where the coefficients $c_{k,1}$, $c_{k,2}$, $c_{k,3}$ and $c_{k,3}$ are respectively given by \vspace{-0.05cm}
\begin{equation*} \vspace{-0.2cm}
  \hspace{-1cm} c_{k,1}  = \left( \rho ^2\left( \mu _k+\delta +1 \right) ^2+\left( 1-\rho ^2 \right) \delta ^2\mu _k+\delta ^2 \right) N^2
\end{equation*}
\begin{equation*} \vspace{-0.2cm}
  + \Big( \left( \left( 2\mu _k+3\delta +2-\delta \mu _k \right) \rho ^2+\left( 1+\mu _k \right) \delta \right) \delta \mu _k\left| f_k \right|^2 ,
\end{equation*}
\begin{equation*} \vspace{-0.2cm}
  +\left( \mu _k+\delta +2 \right) ^2-\rho ^2\left( \mu _k+\delta +1 \right) ^2-2\rho ^2\delta \mu _k-2 \Big) N
\end{equation*}
\begin{equation}\label{ckk_1} \vspace{-0.1cm}
+\rho ^2\delta ^2\mu _{k}^{2}\left| f_k \right|^4+2\left( \left( 1-\rho ^2 \right) \left( \mu _k+\delta \right) +2 \right) \delta \mu _k\left| f_k \right|^2 ,
\end{equation}
\begin{equation*}\vspace{-0.2cm}
  c_{k,2} = \left( \left( 1 \!-\! \iota ^2 \right) \left( \mu _k+\delta +1 \right) ^2-\delta \mu _k\left( \mu _k+1+\delta -\iota ^2\delta \right) \right) \rho ^2
\end{equation*}
\begin{equation}\label{ckk_2} \vspace{-0.1cm}
  +\left( \delta +\mu _k+1 \right) \delta \mu _k+\left( \mu _k+\delta +1 \right) ^2-\left( \mu _k+1 \right) \iota ^2\delta ^2,
\end{equation}
\begin{equation*} \vspace{-0.2cm}
  \hspace{-0.9cm} c_{k,3}= \Big( \left( \left( 3-2\iota ^2 \right) \left( \mu _k+1 \right) +\left( \iota ^2-1 \right) \delta \left( \mu _k-3 \right) \right) \rho ^2
\end{equation*}
\begin{equation*} \vspace{-0.2cm}
  \hspace{-0.9cm} +\left( \delta +1-\iota ^2\delta \right) \left( \mu _k+1 \right) \Big) \delta \mu _k\left| f_k \right|^2
\end{equation*}
\begin{equation*} \vspace{-0.2cm}
  \hspace{-1.65cm} +\left( \left( \iota ^2-1 \right) \left( \mu _k+\delta +1 \right) ^2+\left( \iota ^2-2 \right) 2\mu _k\delta \right) \rho ^2
\end{equation*}
\begin{equation}\label{ckk_3} \vspace{-0.1cm}
  +\left( 1-\iota ^2 \right) \left( \mu _k+\delta +2 \right) ^2+2\mu _k\delta +2\mu _k+2\delta -1-2\iota ^2 ,
\end{equation}
\begin{equation*} \vspace{-0.05cm}
  \hspace{-2.7cm} c_{k,4}=\left( 1-\iota ^2 \right) \rho ^2\delta ^2\mu _{k}^{2}\left| f_k \right|^4 + 2 \delta \mu _k\left| f_k \right|^2
\end{equation*}
\begin{equation}\label{ckk_4}
  \times \left( \left( 1-\iota ^2 \right) \left( 1-\rho ^2 \right) \left( \mu _k+\delta +1 \right) +\left( 2-\iota ^2 \right) \left( 1+\rho ^2 \right) \right)
\end{equation}

\begin{figure*}[b]
\hrulefill 
\setcounter{equation}{73}
\begin{equation*} \vspace{-0.2cm}
  \mathrm{E}\left\{ \left| \omega _{ki}^{1} \right|^2 \right\} = \kappa ^2M \Big( \left( \left( \left( 1 \!-\!\rho ^2 \right) \left( \delta N+2 \right) +\rho ^2\delta \left| f_i \right|^2 \right) \delta \left| f_k \right|^2+\left( 1 \!-\! \rho ^2 \right) N+\rho ^2\left| \mathbf{\bar{h}}_{k}^{H}\mathbf{\bar{h}}_i \right|^2+2\rho ^2\delta \mathrm{Re}\left( f_{k}^{\ast}f_i\mathbf{\bar{h}}_{i}^{H}\mathbf{\bar{h}}_k \right) \right) \iota ^2\mu _k\mu _iM
\end{equation*}
\begin{equation*} \vspace{-0.2cm}
  +\left( \left( 1-\rho ^2 \right) \delta +1 \right) \mu _k\mu _iN^2 +\left( \left( 1-\iota ^2 \right) \left( 1-\rho ^2 \right) \left( \delta ^2\left| f_k \right|^2+1 \right) +\delta \left| f_k \right|^2+\rho ^2\delta \left| f_i \right|^2 \right) \mu _k\mu _iN
\end{equation*}
\begin{equation}\label{w1_ki_norm2_value} \vspace{-0.1cm}
  +\left( \rho ^2\delta ^2\left| f_k \right|^2\left| f_i \right|^2+\rho ^2\left| \mathbf{\bar{h}}_{k}^{H}\mathbf{\bar{h}}_i \right|^2+2\delta \left( 1-\rho ^2 \right) \left| f_k \right|^2+2\delta \rho ^2\mathrm{Re}\left( f_{k}^{\ast}f_i\mathbf{\bar{h}}_{i}^{H}\mathbf{\bar{h}}_k \right) \right) \left( 1-\iota ^2 \right) \mu _k\mu _i \Big)
\end{equation}
\begin{equation*} \vspace{-0.2cm}
  \mathrm{E}\left\{ \left| \omega _{ki}^{2} \right|^2 \right\} = \kappa ^2M \Big( \left( \left( \delta N+2 \right) \delta \left| f_k \right|^2+N \right) \iota ^2\mu _kM+\left( \delta +1 \right) \mu _kN^2
\end{equation*}
\begin{equation}\label{w2_ki_norm2_value} \vspace{-0.1cm}
  +\left( \left( 1-\iota ^2 \right) \left( \delta ^2\left| f_k \right|^2+1 \right) +\delta \left| f_k \right|^2 \right) \mu _kN+2\left( 1-\iota ^2 \right) \delta \mu _k\left| f_k \right|^2 \Big)
\end{equation}
\begin{equation*} \vspace{-0.2cm}
  \mathrm{E}\left\{ \left| \omega _{ki}^{3} \right|^2 \right\} =\kappa ^2M \Big( \left( \left( \delta N \!+\! 2 \right) \delta \rho ^2\left| f_i \right|^2+N+\left( 1 \!-\! \rho ^2 \right) \left( \delta N \!+\! 2 \right) \delta N \right) \iota ^2\mu _iM      +\left( \left( \left( 1 \!-\! \iota ^2 \right) \delta +1 \right) \left( 1 \!-\! \rho ^2 \right) \delta +\delta +1 \right) \mu _iN^2
\end{equation*}
\begin{equation}\label{w3_ki_norm2_value} \vspace{-0.1cm}
  +\left( \left( \rho ^2\delta ^2\left| f_i \right|^2 + 2\left( 1-\rho ^2 \right) \delta +1 \right) \left( 1-\iota ^2 \right) +\rho ^2\delta \left| f_i \right|^2 \right) \mu _iN       +2\left( 1-\iota ^2 \right) \rho ^2\delta \mu _i\left| f_i \right|^2  \Big)
\end{equation}
\begin{equation}\label{w4_ki_norm2_value} \vspace{-0.1cm}
  \mathrm{E}\left\{ \left| \omega _{ki}^{4} \right|^2 \right\} =\kappa ^2M\left( \left( \delta ^2N+2\delta +1 \right) \iota ^2NM+\left( \left( 1-\iota ^2 \right) \delta ^2+2\delta +1 \right) N^2+\left( 1+2\delta \right) \left( 1-\iota ^2 \right) N \right)
\end{equation}
\setcounter{equation}{71}
\end{figure*}

Similar to \eqref{hk_hk}, we expand $\mathbf{h}_{k}^{H}\mathbf{\Phi }^H\mathbf{G}^H \bm{\chi} \mathbf{G\Phi \Theta h}_{i}^{}$ as
\begin{equation*} \vspace{-0.1cm}
  \hspace{-1.3cm} \mathbf{h}_{k}^{H}\mathbf{\Phi }^H\mathbf{G}^H \bm{\chi} \mathbf{G\Phi \Theta h}_{i}^{} =  \frac{\beta}{\delta +1}\sqrt{\frac{\alpha _k\alpha _i}{\left( \mu _k+1 \right) \left( \mu _i+1 \right)}}
\end{equation*}
\begin{equation*} \vspace{-0.2cm}
   \hspace{-3.3cm} \times \Big( \; {\underbrace{\sqrt{\mu _k\mu _i}\mathbf{\bar{h}}_{k}^{H}\mathbf{A\Theta \bar{h}}_{i}^{}}_{\omega _{ki}^{1}}}
   + {\underbrace{\sqrt{\mu _k}\mathbf{\bar{h}}_{k}^{H}\mathbf{A\Theta \tilde{h}}_{i}^{}}_{\omega _{ki}^{2}}}
\end{equation*}
\begin{equation}\label{hk_hi} \vspace{-0.1cm}
   \hspace{-3cm} + {\underbrace{\sqrt{\mu _i}\mathbf{\tilde{h}}_{k}^{H}\mathbf{A\Theta \bar{h}}_{i}^{}}_{\omega _{ki}^{3}}}
   + {\underbrace{\mathbf{\tilde{h}}_{k}^{H}\mathbf{A\Theta \tilde{h}}_{i}^{}}_{\omega _{ki}^{4}}} \;  \Big).
\end{equation}
Thus, the expectation $\mathrm{E}\left\{ \left| \mathbf{h}_{k}^{H}\mathbf{\Phi }^H\mathbf{G}^H \bm{\chi} \mathbf{G\Phi \Theta h}_{i}^{} \right|^2 \right\}$ can be expressed as
\begin{equation*} \vspace{-0.2cm}
  \hspace{-4.5cm} \mathrm{E}\left\{ \left| \mathbf{h}_{k}^{H}\mathbf{\Phi }^H\mathbf{G}^H \bm{\chi}\mathbf{G\Phi \Theta h}_{i}^{} \right|^2 \right\}
\end{equation*}
\begin{equation*} \vspace{-0.2cm}
  \hspace{-2.2cm} =\frac{\beta ^2\alpha _k\alpha _i}{\left( \delta +1 \right) ^2\left( \mu _k+1 \right) \left( \mu _i+1 \right)}\mathrm{E}\left\{ \left| \sum_{s=1}^4{\omega _{ki}^{s}} \right|^2 \right\}
\end{equation*}
\begin{equation*} \vspace{-0.2cm}
  \hspace{-4.6cm} =\frac{\beta ^2\alpha _k\alpha _i}{\left( \delta +1 \right) ^2\left( \mu _k+1 \right) \left( \mu _i+1 \right)}
\end{equation*}
\begin{equation*} \vspace{-0.2cm}
  \times \left( \sum_{s=1}^4{\mathrm{E}\left\{ \left| \omega _{ki}^{s} \right|^2 \right\}}+2\sum_{s=1}^4{\sum_{t=s+1}^4{\mathrm{E}\left\{ \mathrm{Re}\left( \omega _{ki}^{s}\left( \omega _{ki}^{t} \right) ^* \right) \right\}}} \right)
\end{equation*}
\begin{equation}\label{hk_hi_norm2_expand} \vspace{-0.1cm}
  \hspace{-1.9cm} \overset{\left( a \right)}{=}\frac{\beta ^2\alpha _k\alpha _i}{\left( \delta +1 \right) ^2\left( \mu _k+1 \right) \left( \mu _i+1 \right)}\times \sum_{s=1}^4{\mathrm{E}\left\{ \left| \omega _{ki}^{s} \right|^2 \right\}}.
\end{equation}
Step $\left( a \right)$ in \eqref{hk_hi_norm2_expand} is obtained by removing the zero terms. Again, similar to the calculation for ${\rm E}\left\{ {{{\left| {\omega _{kk}^1} \right|}^2}} \right\}$, the expectations in \eqref{hk_hi_norm2_expand} are obtained as \eqref{w1_ki_norm2_value}-\eqref{w4_ki_norm2_value} at the bottom of this page.
\addtocounter{equation}{4}
Therefore, the expectation ${\rm E}\left\{ {{{\left| {{\bf{h}}_k^H{{\bf{\Phi }}^H}{{\bf{G}}^H}{\bf{G\Phi \Theta h}}_i^{}} \right|}^2}} \right\}$ can be obtained as \vspace{-0.05cm}
\begin{equation*} \vspace{-0.2cm}
  \mathrm{E}\left\{ \left| \mathbf{h}_{k}^{H}\mathbf{\Phi }^H\mathbf{G}^H \bm{\chi} \mathbf{G\Phi \Theta h}_{i}^{} \right|^2 \right\} = \frac{\kappa ^2\beta ^2\alpha _k\alpha _iM}{\left( \delta +1 \right) ^2\left( \mu _k+1 \right) \left( \mu _i+1 \right)}
\end{equation*}
\begin{equation}\label{hk_hi_norm2} \vspace{-0.05cm}
  \times \left( z_{ki,1}\iota ^2M+z_{ki,2}N^2+z_{ki,3}N+z_{ki,4} \right) \triangleq \gamma _{ki},
\end{equation}
where the coefficients $z_{ki,1}$ - $z_{ki,4}$ are respectively given by
\begin{equation*} \vspace{-0.2cm}
  \hspace{-4.2cm} z_{ki,1} = \left( \mu _i+1-\rho ^2\mu _i \right) \delta ^2N^2
\end{equation*}
\begin{equation*} \vspace{-0.2cm}
  \hspace{-1.55cm} + \Big( \left( \mu _i+1-\rho ^2\mu _i \right) \delta ^2\mu _k\left| f_k \right|^2+\rho ^2\delta ^2\mu _i\left| f_i \right|^2
\end{equation*}
\begin{equation*} \vspace{-0.2cm}
  \hspace{-1.7cm} +\left( \mu _k+2\delta +1 \right) \left( \mu _i+1-\rho ^2\mu _i \right) +\rho ^2\mu _i   \Big) N
\end{equation*}
\begin{equation*} \vspace{-0.2cm}
  +\left( 2\delta \left| f_i \right|^2+\mu _k\left| \mathbf{\bar{h}}_{k}^{H}\mathbf{\bar{h}}_i \right|^2+2\delta \mu _k\mathrm{Re}\left( f_{k}^{\ast}f_i\mathbf{\bar{h}}_{i}^{H}\mathbf{\bar{h}}_k \right) \right) \rho ^2\mu _i
\end{equation*}
\begin{equation}\label{zki_1} \vspace{-0.1cm}
  \hspace{-1.4cm}  +\left( \rho ^2\delta \mu _i\left| f_i \right|^2+2\mu _i\left( 1-\rho ^2 \right) +2 \right) \delta \mu _k\left| f_k \right|^2  ,
\end{equation}
\begin{equation*} \vspace{-0.2cm}
  \hspace{-1.5cm} z_{ki,2}= \left( \left( \delta +1 \right) \mu _k+\left( \delta +1 \right) ^2-\iota ^2\delta ^2 \right) \left( \mu _i+1 \right)
\end{equation*}
\begin{equation}\label{zki_2} \vspace{-0.1cm}
  \hspace{-1.5cm} -\left( \mu _k+\delta +1-\iota ^2\delta \right) \rho ^2\delta \mu _i-1 ,
\end{equation}
\begin{equation*} \vspace{-0.1cm}
  \hspace{-0.4cm} z_{ki,3} =  \left( \left( 1-\iota ^2 \right) \delta \left( \mu _i+1-\rho ^2\mu _i \right) +\mu _i+1 \right) \delta \mu _k\left| f_k \right|^2
\end{equation*}
\begin{equation*} \vspace{-0.1cm}
  +\left( \left( \mu _i+1 \right) \left( \mu _k+2\delta +1 \right) -\left( \mu _k+2\delta \right) \rho ^2\mu _i \right) \left( 1-\iota ^2 \right)
\end{equation*}
\begin{equation}\label{zki_3} \vspace{-0.1cm}
  \hspace{-2.9cm}  +\left( \mu _k+\delta +1-\iota ^2\delta \right) \rho ^2\delta \mu _i\left| f_i \right|^2 ,
\end{equation}
\begin{equation*} \vspace{-0.2cm}
  z_{ki,4} =  \left( \left( \delta \left| f_i \right|^2-2 \right) \rho ^2\mu _i+2\mu _i+2 \right) \left( 1-\iota ^2 \right) \delta \mu _k\left| f_k \right|^2
\end{equation*}
\begin{equation*} \vspace{-0.1cm}
  +\left( 1-\iota ^2 \right) \rho ^2\mu _k\mu _i\left( \left| \mathbf{\bar{h}}_{k}^{H}\mathbf{\bar{h}}_i \right|^2+2\delta \mathrm{Re}\left( f_{k}^{\ast}f_i\mathbf{\bar{h}}_{i}^{H}\mathbf{\bar{h}}_k \right) \right)
\end{equation*}
\begin{equation}\label{zki_4}
  \hspace{-4cm} +2\left( 1-\iota ^2 \right) \rho ^2\delta \mu _i\left| f_i \right|^2 ,
\end{equation}

\begin{figure*}[t]
\setcounter{equation}{90}
\begin{align}\label{zeta_E1}
  & \mathrm{E}\left\{ \mathbf{h}_{s}^{H}\mathbf{\Theta }^H\mathbf{\Phi }^H\mathbf{G}^H\bm{\chi }^H\left( \mathbf{1}_{M,m} \right) ^H\mathbf{1}_{M,m} \bm{\chi} \mathbf{ G\Phi \Theta h}_s \right\} =\mathrm{E}\left\{ \mathbf{h}_{s}^{H}\mathbf{\Theta }^H\mathbf{\Phi }^H\mathrm{E}\left\{ \mathbf{G}^H\bm{\chi }^H\left( \mathbf{1}_{M,m} \right) ^H\mathbf{1}_{M,m} \bm{\chi} \mathbf{G} \right\} \mathbf{\Phi \Theta h}_s \right\}   \nonumber \\
  & =\frac{\kappa ^2\beta}{\delta +1}\mathrm{E}\left\{ \delta \mathbf{h}_{s}^{H}\mathbf{\Theta }^H\mathbf{\Phi }^H\mathbf{a}_{N}^{}\left( \phi _{t}^{a},\phi _{t}^{e} \right) \mathbf{a}_{N}^{H}\left( \phi _{t}^{a},\phi _{t}^{e} \right) \mathbf{\Phi \Theta h}_s \right\} +\frac{\beta}{\delta +1}\mathrm{E}\left\{ \mathbf{h}_{s}^{H}\mathbf{h}_s \right\}   \nonumber \\
  & =\frac{\kappa ^2\delta \beta \alpha _s}{\left( \delta +1 \right) \left( \mu _s+1 \right)}\left( \mu _s\mathrm{E}\left\{ \left| \mathbf{a}_{N}^{H}\left( \phi _{t}^{a},\phi _{t}^{e} \right) \mathbf{\Phi \Theta \bar{h}}_{s}^{} \right|^2 \right\} +\mathrm{E}\left\{ \left| \mathbf{a}_{N}^{H}\left( \phi _{t}^{a},\phi _{t}^{e} \right) \mathbf{\Phi \Theta \tilde{h}}_{s}^{} \right|^2 \right\} \right) +\frac{\kappa ^2\beta \alpha _sN}{\delta +1} \nonumber \\
  & \overset{\left( a \right)}{=}\frac{\kappa ^2\beta \alpha _s}{\left( \delta +1 \right) \left( \mu _s+1 \right)}\left( \rho ^2\delta \mu _s\left| f_s \right|^2+\left( \left( 1-\rho ^2 \right) \delta \mu _s+\delta +\mu _s+1 \right) N \right)
\end{align}
\hrulefill 
\setcounter{equation}{93}
\begin{equation}\label{zeta_value}
  \mathrm{E}\left\{ \left| \mathbf{1}_{M,m}\bm{\chi} \mathbf{ G\Phi \Theta HPx} \right|^2 \right\} =\sum_{s=1}^K{\frac{\kappa ^2p_s\beta \alpha _s}{\left( \delta +1 \right) \left( \mu _s+1 \right)}\left( \rho ^2\delta \mu _s\left| f_s \right|^2+\left( \left( 1-\rho ^2 \right) \delta \mu _s+\delta +\mu _s+1 \right) N \right)}\triangleq \zeta
\end{equation}
\hrulefill 
\setcounter{equation}{82}
\vspace{-0.2cm}
\end{figure*}

Then, we focus on the expectations $\mathrm{E}\left\{ \mathrm{DN}_k \right\}$, $\mathrm{E}\left\{ \mathrm{AN}_k \right\}$ and $\mathrm{E}\left\{ \mathrm{QN}_k \right\}$.
According to \eqref{DN_k}-\eqref{QN_k}, they can be expanded respectively as \vspace{-0.1cm}
\begin{equation}\label{EDNk_expand} \vspace{-0.15cm}
  \mathrm{E}\left\{ \mathrm{DN}_k \right\} =\tau ^2\sigma _{\mathrm{RF}}^{2}\sum_{m=1}^M{\mathrm{E}\left\{ \left| \mathbf{h}_{k}^{H}\mathbf{\Phi }^H\mathbf{G}^H\mathbf{1}_{M,m}^{H} \right|^2 \right\}} ,
\end{equation}
\begin{equation}\label{EANk_expand} \vspace{-0.15cm}
  \mathrm{E}\left\{ \mathrm{AN}_k \right\} =\tau ^2\sigma ^2\sum_{m=1}^M{\mathrm{E}\left\{ \left| \mathbf{h}_{k}^{H}\mathbf{\Phi }^H\mathbf{G}^H\mathbf{1}_{M,m}^{H} \right|^2 \right\}} ,
\end{equation}
\begin{equation}\label{EQNk_expand} \vspace{-0.1cm}
  \mathrm{E} \left\{ \mathrm{QN}_k \right\} =\tau \left( 1 \!-\! \tau \right) \sum_{m=1}^M{\left[ \mathbf{S}_Q \right] _{mm}\mathrm{E}\left\{ \left| \mathbf{h}_{k}^{H}\mathbf{\Phi }^H\mathbf{G}^H\mathbf{1}_{M,m}^{H} \right|^2 \right\}} .
\end{equation}
To obtain the expectations in \eqref{EDNk_expand} and \eqref{EANk_expand}, we calculate the value of $\mathrm{E}\left\{ \left| \mathbf{h}_{k}^{H}\mathbf{\Phi }^H\mathbf{G}^H\mathbf{1}_{M,m}^{H} \right|^2 \right\}$, which is shown as follows:
\begin{equation*} \vspace{-0.1cm}
  \hspace{-5.2cm} \mathrm{E}\left\{ \left| \mathbf{h}_{k}^{H}\mathbf{\Phi }^H\mathbf{G}^H\mathbf{1}_{M,m}^{H} \right|^2 \right\}
\end{equation*}
\begin{equation*} \vspace{-0.1cm}
  \hspace{-2.6cm} =\mathrm{E}\left\{ \mathbf{h}_{k}^{H}\mathbf{\Phi }^H\mathrm{E}\left\{ \mathbf{G}^H\mathbf{1}_{M,m}^{H}\mathbf{1}_{M,m}^{}\mathbf{G} \right\} \mathbf{\Phi h}_{k}^{} \right\}
\end{equation*}
\begin{equation*} \vspace{-0.2cm}
  \overset{\left( a \right)}{=}\mathrm{E}\left\{ \mathbf{h}_{k}^{H}\mathbf{\Phi }^H\frac{\beta}{\delta +1}\left( \delta \mathbf{a}_{N}^{}\left( \phi _{t}^{a},\phi _{t}^{e} \right) \mathbf{a}_{N}^{H}\left( \phi _{t}^{a},\phi _{t}^{e} \right) +\mathbf{I}_M \right) \mathbf{\Phi h}_k \right\}
\end{equation*}
\begin{equation*} \vspace{-0.2cm}
  \hspace{-2.05cm} \overset{\left( b \right)}{=}\frac{\beta \alpha _k}{\left( \delta +1 \right) \left( \mu _k+1 \right)}\delta \mathrm{E}\left\{ \left| \mathbf{a}_{N}^{H}\left( \phi _{t}^{a},\phi _{t}^{e} \right) \mathbf{\Phi \tilde{h}}_{k}^{} \right|^2 \right\}
\end{equation*}
\begin{equation*} \vspace{-0.1cm}
  \hspace{-1.5cm} +\frac{\beta \alpha _k}{\left( \delta +1 \right) \left( \mu _k+1 \right)}\left( \delta \mu _k\left| f_k \right|^2+\left( \mu _k+1 \right) N \right)
\end{equation*}
\begin{equation}\label{bar_w_k}
  \hspace{-0.3cm} \overset{\left( c \right)}{=}\frac{\beta \alpha _k}{\left( \delta +1 \right) \left( \mu _k+1 \right)}\left( \delta \mu _k\left| f_k \right|^2+\left( \mu _k+\delta +1 \right) N \right) \triangleq \varpi _k,
\end{equation}
where steps $\left( a \right)$ and $\left( b \right)$ are obtained by substituting \eqref{channel_G} and \eqref{channel_H} into the derivation respectively and then removing the zero terms. Step $\left( c \right)$ is based on the calculation
\begin{equation*}
  \mathrm{E} \! \left\{ \left| \mathbf{a}_{N}^{H}\left( \phi _{t}^{a},\phi _{t}^{e} \right) \mathbf{\Phi \tilde{h}}_{k}^{} \right|^2 \right\} \!=\! \mathrm{E}\! \left\{ \left| \sum_{n=1}^N{a_{Nn}^{\ast}\left( \phi _{t}^{a},\phi _{t}^{e} \right) e^{j\theta _n}\tilde{h}_{nk}^{}} \right|^2 \right\}
\end{equation*}
\begin{equation}\label{Efor_DNAN}
  = \! \mathrm{E} \! \left\{ \sum_{n=1}^N{\left| a_{Nn}^{\ast}\left( \phi _{t}^{a},\phi _{t}^{e} \right) e^{j\theta _n}\tilde{h}_{nk}^{} \right|^2} \right\} \!=\! \mathrm{E} \! \left\{ \sum_{n=1}^N{\left| \tilde{h}_{nk}^{} \right|^2} \right\} \!=\! N.
\end{equation}
Thus, we have
\begin{equation}\label{EDNk_value} \vspace{-0.01cm}
  \mathrm{E}\left\{ \mathrm{DN}_k \right\} = \tau ^2\sigma _{\mathrm{RF}}^{2}\varpi _kM ,
\end{equation}
\begin{equation}\label{EANk_value} \vspace{-0.05cm}
  \mathrm{E}\left\{ \mathrm{AN}_k \right\} = \tau ^2\sigma ^2\varpi _kM.
\end{equation}

Furthermore, to obtain the expectation in \eqref{EQNk_expand}, we need additionally derive the expectation $\mathrm{E}\left\{ \left| \mathbf{1}_{M,m}\bm{\chi} \mathbf{G\Phi \Theta HPx} \right|^2 \right\}$ in $\left[ \mathbf{S}_{\mathrm{Q}} \right] _{mm}$ in \eqref{S_Q_mm}, which can be further expressed as
\begin{equation*}
  \hspace{-4.9cm} \mathrm{E}\left\{ \left| \mathbf{1}_{M,m} \bm{\chi} \mathbf{G\Phi \Theta HPx} \right|^2 \right\}
\end{equation*}
\begin{equation*}
  \hspace{-0.2cm} =\mathrm{E}\left\{ \mathbf{1}_{M,m} \bm{\chi} \mathbf{G\Phi \Theta HPP}^H\mathbf{H}^H\mathbf{\Theta }^H\mathbf{\Phi }^H\mathbf{G}^H\bm{\chi }^H\left( \mathbf{1}_{M,m} \right) ^H \right\}
\end{equation*}
\begin{equation*}
  = \!\mathrm{tr} \!\left( \mathbf{PP}^{\!H}\mathrm{E} \! \left\{ \mathbf{H}^{\!H}\mathbf{\Theta }^{\!H}\mathbf{\Phi }^{\!H}\mathbf{G}^{\!H}\bm{\chi }^{H} \! \left( \mathbf{1}_{M,m} \right) ^{\!H} \!\mathbf{1}_{M,m} \bm{\chi} \mathbf{G\Phi \Theta H} \right\} \right)
\end{equation*}
\begin{equation}\label{zeta_expand}
  \hspace{-0.2cm} =\sum_{s=1}^K{p_s\mathrm{E}  \left\{ \mathbf{h}_{s}^{H}\mathbf{\Theta }^{H}\mathbf{\Phi }^{H}\mathbf{G}^{H}\bm{\chi }^H \! \left( \mathbf{1}_{M,m} \right) \! ^H\mathbf{1}_{M,m} \bm{\chi} \mathbf{ G\Phi \Theta h}_s \right\}}
\end{equation}
where ${\bf{1}}_{M,m} \in \mathbb{C}^{1 \times M}$ is the vector whose m\textit{th} element is 1, while the rest elements are zero.
The last expectation in \eqref{zeta_expand} can be calculated as \eqref{zeta_E1} at the top of this page,
\addtocounter{equation}{1}
where step $\left( a \right)$ is based on the following calculations:
\begin{equation*} \vspace{-0.2cm}
  \hspace{-1.2cm} \mathrm{E}\left\{ \left| \mathbf{a}_{N}^{H}\left( \phi _{t}^{a},\phi _{t}^{e} \right) \mathbf{\Phi \Theta \bar{h}}_{s}^{} \right|^2 \right\} =\mathrm{E}\left\{ \left| \sum_{n=1}^N{e^{j\varepsilon _n}f_{s,n}} \right|^2 \right\}
\end{equation*}
\begin{equation*} \vspace{-0.2cm}
  =\mathrm{E}\left\{ \sum_{n=1}^N{\left| e^{j\varepsilon _n}f_{s,n} \right|^2}+\sum_{n_1=1}^N{\sum_{n_2\ne n_1}^N{e^{j\varepsilon _{n_1}}f_{s,n_1}e^{-j\varepsilon _{n_2}}f_{s,n_2}^{\ast}}} \right\}
\end{equation*}
\begin{equation*} \vspace{-0.2cm}
  \hspace{-4.1cm} =N+\rho ^2\sum_{n_1=1}^N{\sum_{n_2\ne n_1}^N{f_{s,n_1}f_{s,n_2}^{\ast}}}
\end{equation*}
\begin{equation}\label{zeta_E2} \vspace{-0.1cm}
  \hspace{-1.5cm} =N+\rho ^2\left( \left| f_k \right|^2-N \right) =\left( 1-\rho ^2 \right) N+\rho ^2\left| f_k \right|^2  ,
\end{equation}
\begin{equation}\label{zeta_E3} \vspace{-0.1cm}
  \mathrm{E}\left\{ \left| \mathbf{a}_{N}^{H}\left( \phi _{t}^{a},\phi _{t}^{e} \right) \mathbf{\Phi \Theta \tilde{h}}_{s}^{} \right|^2 \right\} =\mathrm{E}\left\{ \sum_{n=1}^N{\left| \tilde{h}_{ns}^{} \right|^2} \right\} =N .
\end{equation}
By using \eqref{zeta_expand} and \eqref{zeta_E1}, we arrive at \eqref{zeta_value} at the top of this page.
\addtocounter{equation}{1}
Therefore, from \eqref{S_Q_mm}, \eqref{bar_w_k} and \eqref{zeta_value}, we can obtain \vspace{-0.2cm}
\begin{equation}\label{EQNk_value} \vspace{-0.1cm}
  \mathrm{E}\left\{ \mathrm{QN}_k \right\} = \tau \left( 1-\tau \right) \varpi _k\left( \zeta +\sigma _{\mathrm{RF}}^{2}+\sigma ^2 \right) M .
\end{equation}

Finally, substituting \eqref{hk_hk_norm2}, \eqref{hk_hi_norm2}, \eqref{EDNk_value}, \eqref{EANk_value} and \eqref{EQNk_value} into \eqref{EAR_user_k_approx_afterlemma1}, we obtain \eqref{EAR_user_k_approx} and complete the proof.

\end{appendices}

\bibliographystyle{IEEEtran}
\bibliography{IEEEabrv,Reference}

\begin{IEEEbiography}[{\includegraphics[width=1in,height=1.25in,clip,keepaspectratio]{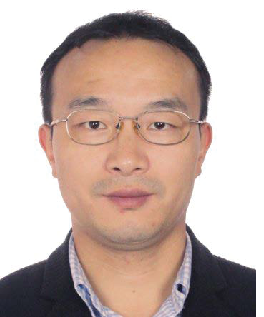}}]{Zhangjie Peng}
received the B.S. degree from Southwest Jiaotong University, Chengdu, China, in 2004, and the M.S. and Ph.D. degrees from Southeast University, Southeast University, Nanjing, China, in 2007, and 2016, respectively, all in Communication and Information Engineering. He is currently an Associate Professor at the College of Information, Mechanical and Electrical Engineering, Shanghai Normal University, Shanghai 200234, China.

His research interests include reconfigurable intelligent surface (RIS), cooperative communications, information theory, physical layer security, and machine learning for wireless communications.
\end{IEEEbiography}

\begin{IEEEbiography}[{\includegraphics[width=1in,height=1.25in,clip,keepaspectratio]{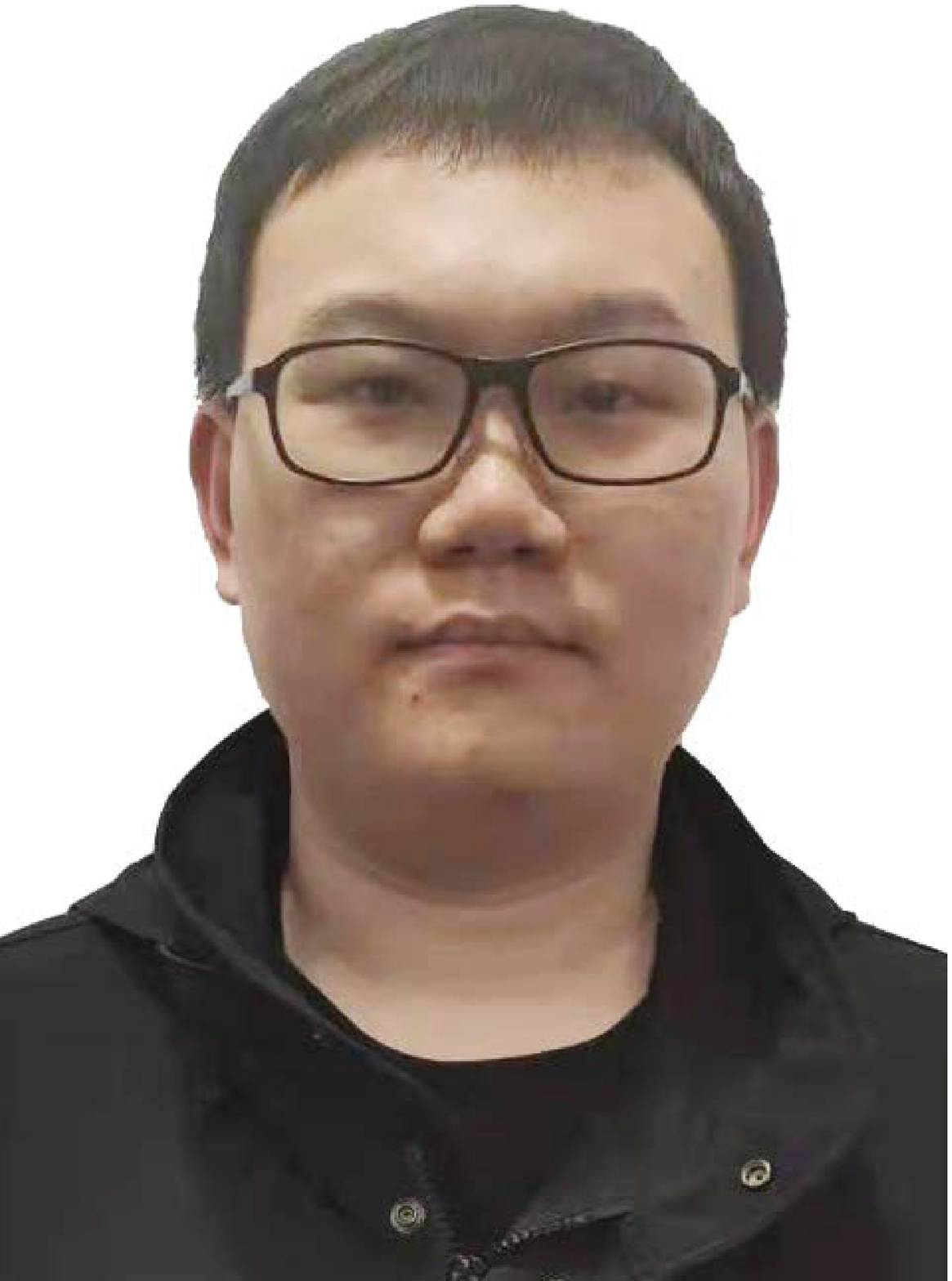}}]{Xianzhe Chen}
received the B.E. degree in information engineering from Zhejiang University, Zhejiang, China, in 2019, and the M.E. degree in information and communication engineering from Shanghai Normal University, Shanghai, China, in 2022. He is currently pursuing the Ph.D. degree with the Department of Electrical and Computer Engineering, The University of British Columbia, Vancouver, Canada.

His major research interests include reconfigurable intelligent surfaces, massive multiple-input multiple-output systems and relaying communications.
\end{IEEEbiography}

\begin{IEEEbiography}[{\includegraphics[width=1in,height=1.25in,clip,keepaspectratio]{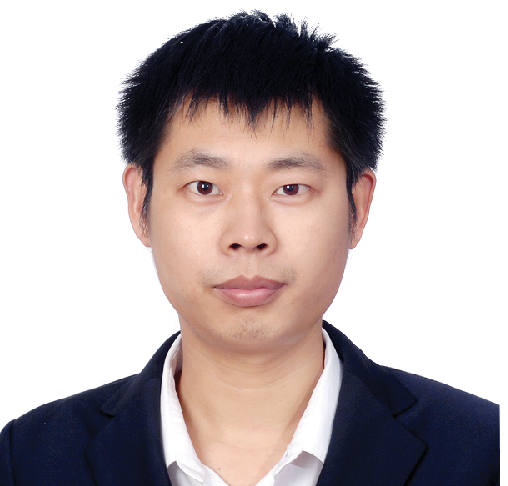}}]{Cunhua Pan}
received the B.S. and Ph.D. degrees from the School of Information Science and Engineering, Southeast University, Nanjing, China, in 2010 and 2015, respectively. From 2015 to 2016, he was a Research Associate at the University of Kent, U.K. He held a post-doctoral position at Queen Mary University of London, U.K., from 2016 and 2019.From 2019 to 2021, he was a Lecturer in the same university. From 2021, he is a full professor in Southeast University.

His research interests mainly include reconfigurable intelligent surfaces (RIS), intelligent reflection surface (IRS), ultra-reliable low latency communication (URLLC) , machine learning, UAV, Internet of Things, and mobile edge computing. He has published over 120 IEEE journal papers. He is currently an Editor of IEEE Wireless Communication Letters, IEEE Communications Letters and IEEE ACCESS. He serves as the guest editor for IEEE Journal on Selected Areas in Communications on the special issue on xURLLC in 6G: Next Generation Ultra-Reliable and Low-Latency Communications. He also serves as a leading guest editor of IEEE Journal of Selected Topics in Signal Processing (JSTSP)  Special Issue on Advanced Signal Processing for Reconfigurable Intelligent Surface-aided 6G Networks, leading guest editor of IEEE Vehicular Technology Magazine on the special issue on Backscatter and Reconfigurable Intelligent Surface Empowered Wireless Communications in 6G, leading guest editor of IEEE Open Journal of Vehicular Technology on the special issue of Reconfigurable Intelligent Surface Empowered Wireless Communications in 6G and Beyond, and leading guest editor of IEEE ACCESS Special Issue on Reconfigurable Intelligent Surface Aided Communications for 6G and Beyond. He is Workshop organizer in IEEE ICCC 2021 on the topic of Reconfigurable Intelligent Surfaces for Next Generation Wireless Communications (RIS for 6G Networks), and workshop organizer in IEEE Globecom 2021 on the topic of Reconfigurable Intelligent Surfaces for future wireless communications. He is currently the Workshops and Symposia officer for Reconfigurable Intelligent Surfaces Emerging Technology Initiative. He is workshop chair for IEEE WCNC 2024, and TPC co-chair for IEEE ICCT 2022. He serves as a TPC member for numerous conferences, such as ICC and GLOBECOM, and the Student Travel Grant Chair for ICC 2019.  He received the IEEE ComSoc Leonard G. Abraham Prize in 2022.
\end{IEEEbiography}

\begin{IEEEbiography}[{\includegraphics[width=1in,height=1.25in,clip,keepaspectratio]{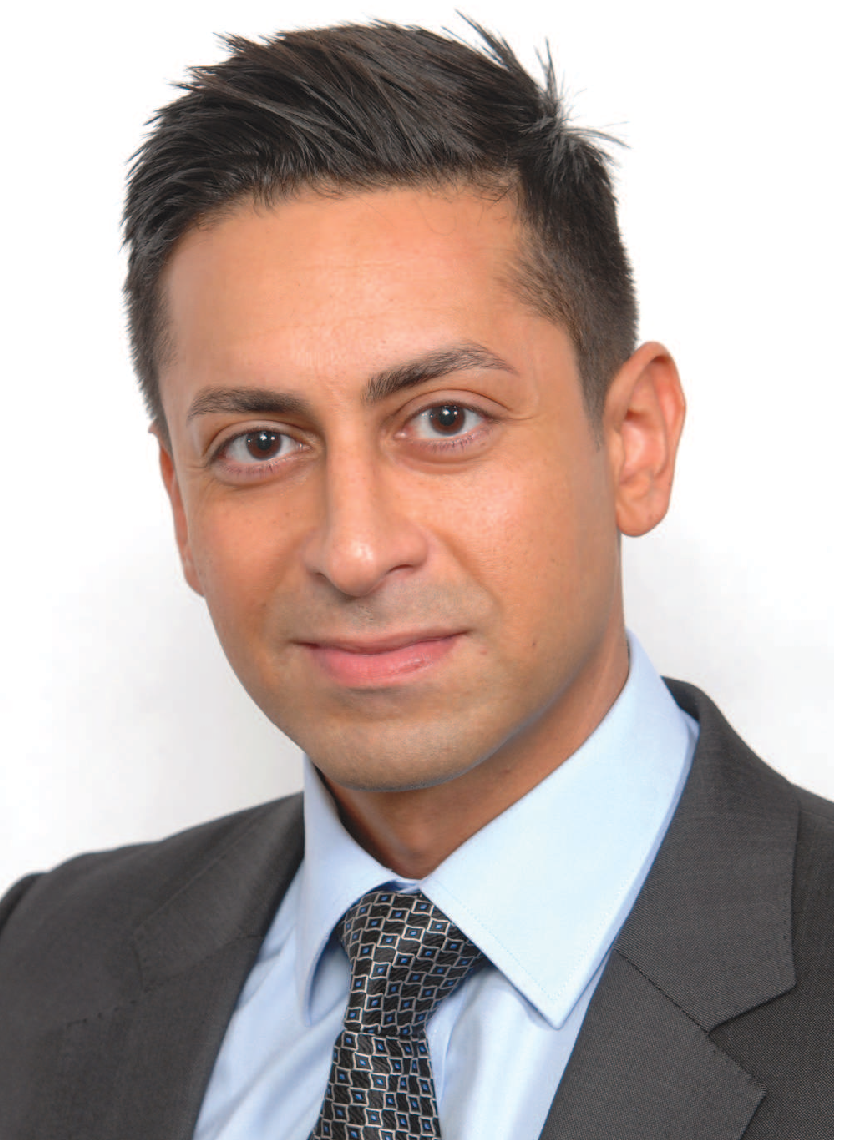}}]{Maged Elkashlan}
received the PhD degree in Electrical Engineering from the University of British Columbia in 2006. From 2007 to 2011, he was with the Commonwealth Scientific and Industrial Research Organization (CSIRO) Australia. During this time, he held visiting faculty appointments at University of New South Wales, University of Sydney, and University of Technology Sydney. In 2011, he joined the School of Electronic Engineering and Computer Science at Queen Mary University of London. He also holds a visiting faculty appointment at Beijing University of Posts and Telecommunications. His research interests fall into the broad areas of communication theory and signal processing.

Dr. Elkashlan is an Editor of the IEEE TRANSACTIONS ON VEHICULAR TECHNOLOGY and the IEEE TRANSACTIONS ON MOLECULAR, BIOLOGICAL AND MULTI-SCALE COMMUNICATIONS. He was an Editor of the IEEE TRANSACTIONS ON WIRELESS COMMUNICATIONS from 2013 to 2018 and the IEEE COMMUNICATIONS LETTERS from 2012 to 2016. He received numerous awards, including the 2022 IEEE Communications Society Leonard G. Abraham Prize, and best paper awards at the 2014 and 2016 IEEE ICC, 2014 CHINACOM, and 2013 IEEE VTC-Spring.
\end{IEEEbiography}

\begin{IEEEbiography}[{\includegraphics[width=1in,height=1.25in,clip,keepaspectratio]{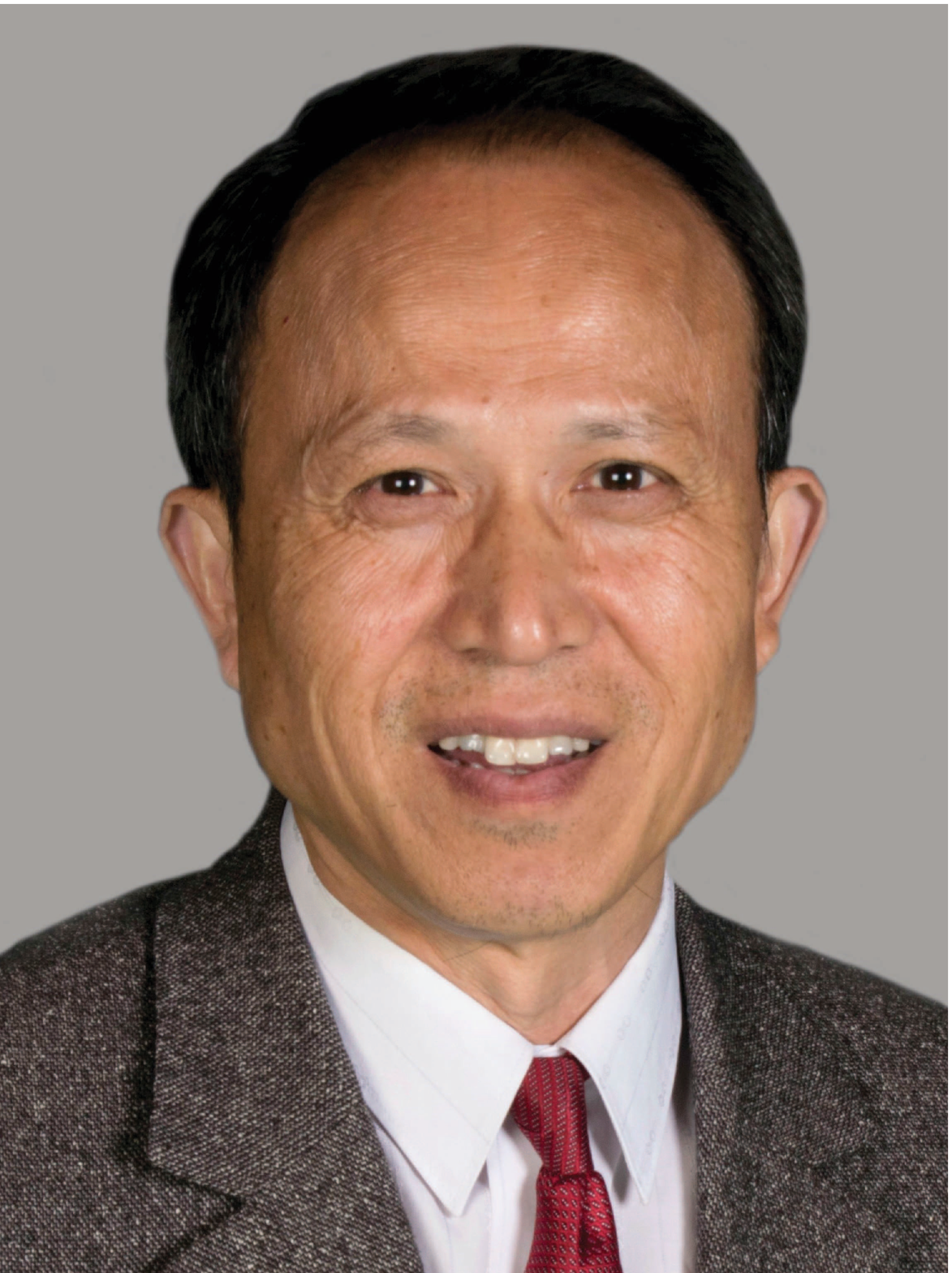}}]{Jiangzhou Wang}
(Fellow, IEEE) is a Professor at the University of Kent, U.K. His research interest is in mobile communications. He has published over 400 papers and 4 books. He was a recipient of the 2022 IEEE Communications Society Leonard G. Abraham Prize and the 2012 IEEE Globecom Best Paper Award. Professor Wang is a Fellow of the Royal Academy of Engineering, U.K., Fellow of the IEEE, and Fellow of the IET. He was the Technical Program Chair of the 2019 IEEE International Conference on Communications (ICC2019), Shanghai, the Executive Chair of the IEEE ICC2015, London, and the Technical Program Chair of the IEEE WCNC2013. He was an IEEE Distinguished Lecturer from 2013 to 2014. He has served as an Editor for a number of international journals, including IEEE Transactions on Communications from 1998 to 2013.
\end{IEEEbiography}

\end{document}